\documentclass[a4paper,reqno,12pt]{article}

\hsize \textwidth
\advance \hsize by -\marginparwidth
\evensidemargin \oddsidemargin
\usepackage{amssymb}
\advance\hoffset by 5mm

\usepackage[titletoc,title]{appendix}

\usepackage[hidelinks]{hyperref}
\usepackage{mathtools}
\usepackage{physics}
\usepackage{dsfont}
\usepackage{bbm}
\usepackage{amsmath}
\usepackage{mathrsfs}
\usepackage{amsthm}
\usepackage{xcolor}

\usepackage{amssymb}
 \usepackage{color}
\usepackage{amsfonts}
\usepackage{amsthm}
\usepackage{amscd}
\usepackage[latin2]{inputenc}
\usepackage{t1enc}
\usepackage[mathscr]{eucal}
\usepackage{indentfirst}
\usepackage{graphicx}
\usepackage{graphics}
\usepackage{pict2e}
\usepackage{epic}

\numberwithin{equation}{section}
\newcommand{\blue}[1]{\textcolor{blue}{#1}}

\newtheorem{theorem}{Theorem}[section]
\newtheorem{corollary}{Corollary}[theorem]
\newtheorem{lemma}[theorem]{Lemma}

\newtheorem{thm}{Theorem}[section]

\newtheorem{prop}[thm]{Proposition}

\newtheorem{rem}[thm]{Remark}

\newcommand\cA{{\mathcal A}}

\newcommand\cF{{\mathcal F}}
\newcommand\cG{{\mathcal G}}
\newcommand\cH{{\mathcal H}}

\newcommand\cM{{\mathcal M}}

\newcommand\cS{{\mathcal S}}

\newcommand\cQ{{\mathcal Q}}
\newcommand\cR{{\mathcal R}}

\newcommand\bB{{\mathbb B}}
\newcommand\bC{{\mathbb C}}
\newcommand\bE{{\mathbb E}}
\newcommand\bN{{\mathbb N}}
\newcommand\bP{{\mathbb P}}
\newcommand\bR{{\mathbb R}}
\newcommand\bT{{\mathbb T}}
\newcommand\bZ{{\mathbb Z}}

\newcommand\bS{{\mathbb S}}

\newcommand\bTd{{\widetilde{\mathbb T}}}

\newcommand\cn{{\mathfrak{cn}}}
\newcommand\sn{{\mathfrak{sn}}}
\newcommand\dn{{\mathfrak{dn}}}

\begin{document}

\title{A stochastic thermalization
  of the Discrete Nonlinear Schr\"{o}dinger Equation }

\author{Amirali Hannani$^1$ and Stefano Olla$^{1,2,3}$
 \thanks{$^1$CEREMADE, UMR CNRS,
 Universit\'e Paris-Dauphine, PSL Research University,\newline
 $^2$Institute Universitaire de France, $^3$GSSI.
 {\tt hannani@ceremade.dauphine.fr}, {\tt olla@ceremade.dauphine.fr}.
 \newline
 This work was partially supported by ANR-15-CE40-0020-01 grant LSD.
 We thank Sourav Chatterjee for stimulating discussions at the early stage of this work.
We also thank Nikolay Tzvetkov for suggesting reference \cite{pava} for the periodic NLSE.}}

% address
% {Stefano Olla (\emph{corresponding author}):
%   CEREMADE, UMR CNRS,\\
% Universit\'e Paris-Dauphine, PSL Research University\\
% 75016 Paris, France\\
% and GSSI, 67100 L'Aquila, Italy}
% \email
% {\tt olla@ceremade.dauphine.fr}

\date{\today}

    \maketitle    
    
    \abstract{We introduce a mass conserving stochastic perturbation of the discrete
      nonlinear Schr\"{o}dinger equation that models
      the action of a heat bath at a given temperature. We prove that the
      corresponding canonical Gibbs distribution is the unique invariant measure.
      In the one-dimensional cubic focusing case {on the torus}, 
      we prove that in the limit for large time,
      continuous approximation, and low temperature,
      the solution converges to the steady wave of the continuous equation
      that minimizes the energy for a given mass.}
    
    \section{Introduction}
\label{sec:introduction}

	Consider the Nonlinear Schr\"{o}dinger Equation in $d$ space dimension:
  	\begin{equation} \label{NLS0}
		\begin{split}  		
                 & i\partial_t \psi(x,t) = -\Delta \psi(x,t) + \kappa  |\psi|^{p-1} \psi(x,t); \qquad
                  p>1,\\
  		& \psi: \Omega^d\times \bR_+ \to \bC; \quad \psi(x,0):= \psi_0(x),
		\end{split}  	
  	\end{equation}
	where $\Omega = \bR$, or $\Omega = \mathbb T_L^1$,
        the circle of length $L$, for the periodic boundary conditions case,
        % $\Delta$ denotes the Laplacian on $\bR^d$ \footnote{Except this expression
	% we  use $\Delta$ to denote discrete Laplacian.},
      % $p>1$ determines the strength  of the
      %   non-linearity,
        $\kappa=- 1$ corresponds to the focusing case,
        and $\kappa=1$ to the defocusing
        % {\color{red} (For  general introduction
        % regarding NLS cf. \cite{cazenave03}, \cite{bourgain}, \cite{tao06})}
      .
        This equation has many conserved quantities,
        in particular the most important are the energy and the mass: 
        	\begin{equation} \label{conserved-intro}
		\begin{split}
                  \mathcal{H}(\psi)=
      \frac12 \int |\partial_x\psi|^2  dx +  \frac{\kappa}{p+1} \int |\psi|^{p+1} dx, 
		 	\qquad \mathcal{M}(\psi)= \int |\psi|^2 dx.
		 \end{split}
               \end{equation}
               In some particular cases (like for $d=1$ and $p=3$),
               the dynamics is completely integrable.

               We are particularly interested
               in the focusing case $\kappa = -1$, where the non-linearity
               contrast the dispersive  effect of the Laplacian. Notice that, thanks to the
               Gagliardo-Nirenberg inequality (cf. \eqref{GNcontinous}),
               $ \mathcal{H}(\psi)$ is still bounded below if $\mathcal{M}(\psi)$
               is fixed, and $p < 1+ \frac 4d$, also known as mass sub-critical case.
               In the one-dimensional mass sub-critical NLS with periodic boundary conditions
               ($d=1$, $p < 5$, and $\Omega =\mathbb T_L^1$), it has been proven that
               the canonical Gibbs measure at temperature $\beta^{-1}$,
               formally defined as
               \begin{equation}
                 \mathcal Z^{-1} \exp{-\beta  \mathcal{H}(\psi)}
                 \delta\left( \mathcal{M}(\psi)=m\right)\prod_x d\psi(x)
                 \label{eq:gibbscont}
             \end{equation}
             is invariant for the dynamics defined by \eqref{NLS0}.
             {Rigorous definition of \eqref{eq:gibbscont} can be found in \cite{LRS88},
             while its invariance for the dynamics is proven in 
             \cite{Bo94}, see also \cite{bourgain},
              \cite{Mc95}, \cite{McVa971}, \cite{McVa972}, \cite{McVa94}}.
            For $p=3$, $d=1$,	\eqref{NLS0} is completely integrable; {hence} it is
             obvious that \eqref{eq:gibbscont} cannot be ergodic,
             not even conditioned to a value of the energy $\mathcal{H}$
             (i.e., the \emph{microcanonical Gibbs measure})
             as there are other conserved quantities
             beyond energy and mass. 
             A natural question is then how to define a stochastic perturbation of
             \eqref{NLS0} such that  
             acts as a \emph{heat bath} at temperature
             $\beta^{-1}$, and such that the resulting stochastic dynamics has
             \eqref{eq:gibbscont} as the
             unique stationary measure. This implies that
             the only conserved quantity of the dynamics should be the mass
             $\mathcal{M}$.
               
             Formally, one way to define such
             stochastic dynamics is to consider the stochastic partial differential equation
             \begin{equation}
               \label{eq:spde}
               i\partial_t \psi(x,t) = -\Delta \psi(x,t) + \kappa  |\psi|^{p-1} \psi(x,t)
               - \gamma \psi(x,t)\left(i \beta^{-1} -
               \frac{\delta\mathcal H (\psi))}{\delta\theta(x)}\right)
               + \sqrt{2 \gamma \beta^{-1}}\psi(x,t) W(x,t),
             \end{equation}
             where $\theta(x)$ is the phase of $\psi(x)$ ($\psi(x) = |\psi(x)|e^{i\theta(x)}$),
             $W(x,t)$ is the standard space-time white noise, and $\gamma>0$ is
             a parameter that regulates the intensity of the contact with the heat bath.
             { Notice that
             $\frac{\delta\mathcal H (\psi))}{\delta\theta(x)}
             = \mathcal Im[\psi(x)^*\Delta\psi(x)]$,} and that
             \eqref{eq:spde} should be intended in the Ito's sense.
             Consequently, the mass $\mathcal M(\psi)$ is still
             formally conserved by this dynamics.
             The heat bath acts with random but continuous rotations of the phase of
             $\psi(x)$ at each point $x$.             
             Because of the singularity in space of the multiplicative white noise $W$ and
             the non-linearities present in \eqref{eq:spde}, it is very
             hard to give sense to the solution of this equation. There is an extensive literature
             on the NLSE with space correlated multiplicative noise (cf. 
             \cite{ADAD99}, \cite{ADAD06}), but it does not include non linearities like
             $\psi(x)\Delta\psi^*(x)$.
             Additive noises have also been studied
             (cf. \cite{Lebowitz2013}, \cite{Lebowitz2016}, \cite{Lebowitz2019})
             but usually do not conserve the mass, and the corresponding dynamics have
             the Grand Canonical Gibbs measure as stationary.
             
             We introduce instead a space discretization of \eqref{eq:spde}, see
             \eqref{sde1d}, whose solution can be defined globally.
             The infinite temperature version of this stochastic evolution was introduced in
             (cf. \cite{viviane}).
             This is a $n^d$ (complex) dimensional stochastic evolution that conserves the mass, 
             and for any given initial mass, the Gibbs measure on the corresponding
             complex sphere
             defined by \eqref{gibbsmesured},
             discrete analogous of \eqref{eq:gibbscont},
             is well defined and invariant.
             We prove in Section \ref{SDsec} that this Gibbs measure is the unique
             invariant measure, and that the distribution of the process starting
             from an arbirtary initial condition converges exponentially in total variation
             to this stationary measure (cf. Theorem \ref{thinmed} and Proposition
             \ref{thlongtimed}). These results on the ergodicity of the stochastic
             dynamics contained in Section \ref{SDsec}
             are general and are valid for any $d,p>1, \kappa=\pm1, n$
             and more general non-linearity.
             Let us emphasize that,  to the best of our knowledge,
             the novelty  of  this dynamics is
             that it is the first 
              \emph{mass conserving} perturbation 
             of the DNLS (Discrete Non-Linear Schr\"{o}dinger),  such that the canonical Gibbs
             measure is the \emph{unique invariant measure},
             determining the dynamics long-time behavior. We should mention
               that in Section 6 of \cite{Lebowitz2016}, a mass conserving noise
               is proposed such that
             the Canonical Gibbs measure remains invariant by the dynamics. 
             However, this dynamics is not studied, and \cite{Lebowitz2016} mainly
             concerns another dynamics, which does not conserve the mass and converges to 
             the Grand Canonical measure. Moreover, the above-mentioned mass conserving
             dynamics is different from ours; in fact, a straightforward analysis suggests
             that our dynamics is more degenerate.
             
             From Section \ref{sec:mainresult} and after, we concentrate on 
             the one-dimensional focusing
             cubic case with periodic boundary conditions ($d=1, p=3, \kappa =-1$).
             For the continuous model,
             the minimizers of the energy $\mathcal H(\psi)$ under the mass constrain
             $\mathcal M(\psi) = m$ are known explicitely \cite{GLT}. These minimizers,
             that we denote by $Q_{m,L}(x)$,
             are unique up to translations and multiplication by a constant phase.
             To these minimizers correspond a class of standing waves
             $\psi(x,t) = e^{i\omega t}Q_{m,L}(x)$, which are solutions of \eqref{NLS0},
             where the frequency $\omega$ is determined by $m$ and $L$.
             We call \emph{solitons} these ground state standing waves, in analogy
             to the traveling solitary waves of the dynamics in $\mathbb R$.
             If $m\le \frac{\pi^2}{L}$ these solitons are constant in space, while for
             $m > \frac{\pi^2}{L}$  are given by the dnoidal elliptic Jacobi
             functions (cf. Appendix \ref{JEF} for the definition, 
             and Chapter 2,3 of \cite{JEF13}
             for properties of these functions) properly rescaled. 
             These non-trivial solitons
             catch the 0-temperature behavior of the  dynamics.
             The purpose of our work is to show that the solution
             $\psi_n(x,t)$ of the stochastic discrete dynamics, for large time
             $t$, large $n$, and small temperature $\beta^{-1}$,
             is close,
             in an opportune norm,  to the continuous soliton.
             The result is contained in  Theorem \ref{thlimit},
             where it is first taken the limit $t\to\infty$ then $n\to\infty$,
             rescaling the temperature with $n$, i.e. $\beta_n \sim \infty$
             faster than $n$. This is a way to interpret
             the \emph{soliton resolution conjecture} (SRC) in
             the periodic case, where there is no possibility
             for the energy to escape to infinity.
             Intuitively, in the 
             periodic case, our 
             dynamics in the zero temperature limit % , act as a heat bath and
             dissipate 
             the excess of the energy without losing any mass, 
             forcing the system to approach the ground state as 
             $t \to \infty$.
             This mechanism is somehow mimicking the dynamics of DNLS in \cite{CH},
             where energy disperse to infinity via a "radiating" part of the field 
             carrying 
             arbitrarily small mass. 
             {In fact, our dynamics
             is partially motivated by \cite{CH}, where Chatterjee % tries to
             proves a "probabilistic" version of the SRC.
             In particular, in  Theorem 3.1 in \cite{CH} it is proven that
             \emph{almost every} ergodic invariant measure satisflythe SRC
             in the time average sense. % However, lack of ergodicity makes
			% the statement of his result relying on rather abstract measures such that 		
			% passing to "more familiar" functional spaces  seems 
			%  complicated (see Theorem 3.1 of \cite{CH}). 
             Our stochastic dynamics provides the uniqueness of the
             invariant ergodic measure and the time mixing property.  }

             In Theorem \ref{thlimit} the limit for $t\to\infty$ follows
             from the ergodic and time mixing properties of the dynamics
             proven in Section \ref{SDsec}.
             Then we have to prove that the discrete Gibbs measure (finite $n$)
             concentrate fast enough in a small neighborhood of the corresponding
             lowest energy configurations, that we call \emph{discrete solitons},
             who converge to the continuous one as $n\to \infty$.
             This relies on large deviation properties of the discrete Gibbs measure,
             proven in Section \ref{sec:lde}. These large deviations
             estimates are based on some precise large deviations
             of the uniform probability measure $\mu_m^n$ on the complex
             $2n$-dimensional sphere $S_m^n$,
             that we prove in Appendix \ref{sec:some-large-devi}, and the
             discrete version of the Gagliardo-Nirenberg inequality, proven in
             Appendix \ref{secGN}.
             The Gibbs measure has a density
             $\exp(-\beta_n \cH_n)$ with
             respect to the uniform measure $\mu_m^n$. Splitting
             the energy $\cH_n= G_n - V_n$, where $G_n$ is the kinetic part,
            and $V_n$ the potential part \eqref{Ham1},
             one can observe that a "typical" 
             configuration w.r.t $\mu_m^n$ has kinetic energy $G_n \sim n^2$.
             The large deviations estimates in Section \ref{sec:some-large-devi}, in particular 
             Lemma \ref{lemLDgradUB}, combined with Gagliardo-Nirenberg  inequality
             \eqref{GNdiscrete3}
              yields: for $0\leq a <2$, 
             the "entropy factor" behaves as
             $\mu_m^n(G_n \sim n^a) \sim \mu_m^n(\cH_n \sim n^a) \sim 
             e^{-(2-a)n \ln n}$. % ., where the first estimate is a consequence
             % of discrete Gagliardo-Nirenberg inequality
             % \eqref{GNdiscrete3} for $a>0$.
             Therefore, taking into account the 
              Boltzmann factor $\exp(-\beta_n \cH_n)$, we have for  $0\leq a <2$:
             $\mu_{\beta_n,m}^n(\cH_n \sim n^a ) \sim e^{-\beta_n n^a} 
             e^{-(2-a)n \ln n}$.
             { Optimizing this estimate on $a\in [0,2)$,
               if $\beta_n \sim O(1)$, then $a=1$ is the optimal value and
               the Gibbs measure concentrates on 
             rather rough configurations with $|\psi(j)-\psi(j-1)| \sim \frac{1}{\sqrt{n}}$,
             so that $G_n \sim n$. This corresponds to the fact that Wiener measure is 
             concentrated on configurations of H\"{o}lder regularity less than $\frac 12$.
             Instead, if $\beta_n \sim O(n)$ we have that $a=0$ is the optimal value and
             this suggests that 
             $\mu_{\beta_n,m}^n$ to concentrates on smooth
             configurations (i.e., with $|\psi(j)-\psi(j-1)| \sim \frac1n$)
             with $\cH_n \sim O(1)$.}
             Notice that minimal
             energy configurations (the discrete solitons), have energy of order
             one as well. However, this scaling is not enough for this measure to 
             concentrate on a small neighborhood of discrete solitons, and we need to go 
             further. Finally, thanks to large deviation estimate \eqref{MAINlowerbound}, 
             we deduce in Theorem \ref{gibbslimit} that scaling $\beta_n >>n$ is sufficient.
             
              In the last step of the proof, we show in Proposition 
              \ref{propdiscompact} that if $\psi_n$ is a configuration 
              with energy close to $E^0_n(m)$, then its linear interpolation 
              $\bar{\psi}_n$ (see \eqref{linearinterpolation1}) is close 
              to the continuous soliton $Q_{m,L}$
              in $H^1$ norm (up to a translation and multiplication by a phase,
              see \eqref{seminorm}),
              for $n$ sufficiently large. In that regard, first we observe that having 
              energy close to $E_0^n(m)$ means the configuration is smooth $G_n \sim
              O(1)$, thanks to the discrete Gagliardo-Nirenberg inequality.
              Subsequently, since for smooth configurations 
              $\cH_n(\psi_n)$ is close to $\cH(\bar{\psi}_n)$ (See Corollary 
              \eqref{energydisccont}), one can conclude 
              by compactness of the minimizing sequence corresponding to the continuous 
              minimization problem characterizing solitons \eqref{Minimiztion}.
              
             Appendix \ref{app:hor}
             contains the proof of the hypoellipticity of the discrete stochastic dynamics,
             necessary for the proof of the ergodicity of Section \ref{SDsec}.
             Since the real and complex part of our field are somehow
             symmetric in the noise, this makes the proof of  the hypoellipticity more
             complicated than usual, and computing three nested commutators
             is necessary 
             (see \eqref{com3}). % The second ingredient is uniqueness of the  invariant 
             % measure, which is proven in Theorem \ref{thinmed}, thanks to a wise
             % separation of the noise into phase and amplitude part. 

             The Gibbs measure of DNLS have been studied both in
              Mathematics (cf. \cite{CHK}, \cite{CH}) and Physics community (cf. 
              \cite{DNLSbook} and references therein: in particular: 
              \cite{RCKG}, \cite{Ru04}, \cite{Rass04}; See also \cite{Gradenigo2021} ). 
              In the physics community, one usually takes the Kinetic energy 
              with a negative sign and study the measure corresponding to Hamiltonian 
              \eqref{Hamd}, by taking $h=1$.
              Although this regime is substantially different
              from ours, and does not correspond to discretization
              of a continuous profile anymore,  interesting phenomena 
              such as discrete breathers is observed (cf. 
              \cite{fl97}, \cite{We99}).\\
              In mathematics community, we can mention most notably  \cite{CHK}, and 
              \cite{CH} {(cf. \cite{K}, for a review)}. 
              In \cite{CHK}, the Hamiltonian \eqref{Hamd} is considered 
              such that $Nh^2 \to 0$, as $h \to 0$, and $N \to \infty$, where $N$
              denotes the 
              number of particles, and $h$ is the interparticle distance.
              These assumptions
              only seems natural in $d \geq 3$. In this regime, certain phase transition 
              happens: When $\beta m^2 < \theta_c $ the Gibbs measure concentrates
              on configurations such that $\psi_n(j) \sim o(n)$, whereas for 
              $\beta m^2 >\theta_c$ breather-like structures appears, where a single site
              has macroscopic mass.\\
              In \cite{CH}, the model is defined on the box $[0,nh]^d$, such that
               $h \to 0$, $n \to \infty$, with $nh \to \infty$. In this regime, the 
               \emph{microcanonical measure}
               corresponding to energy $E$ concentrates 
               on soliton-like configurations in $\bR^d$. \\
               Comparing our result with \cite{CHK}, and \cite{CH}, we highlight the fact
               that different scaling among the parameters $h$, $n$ leads to substantially
               different phenomena: In \cite{CHK}, $Nh^2 \to 0$ makes the Gradient term 
               negligible and phase transition is a consequence of competition
               among potential energy and mass constraint. In \cite{CH}, $nh \to \infty$,
               kinetic and potential energy becomes comparable; however, the mass 
               per particle goes to zero in the limit, demonstrating the macroscopic
               infinite volume, facilitating escape of the energy to infinity and 
               resulting in soliton like behavior. In contrast, in our case we
               take $n \to \infty$, and $nh=1$, representing the finite macroscopic
               volume, and positive mass per particle in the macroscopic limit. 
               This scaling yields a dominant kinetic 
               energy for typical configurations on the sphere of constant mass,
               {and rescaling $\beta_n$ makes the kinetic and potential
               energy comparable}.
              \\
               In particular, these different scaling change our large deviation 
               estimates \eqref{LD:ub}, and \eqref{MAINlowerbound} comparing to 
               estimates in \cite{CH} (See Section 10 of \cite{CH}).

               \section{Stochastic Dynamics}
               \label{SDsec}

   Fix $n \in \mathbb{N}$, let $\chi = \mathbb{C}^{n^d}$ be the configuration space, 
  and denote a typical element of $\chi$ by $\{ \psi(x) \}_{x \in \bTd _n^d}$, 
  where $\bTd_n=\{1,2,\dots,n\}$ is the discrete Torus of size $n$.
 % where we identified $0 \equiv n$. 
  Equivalently, one can see a function on 
  $\bTd_n^d$, $\psi: \bTd_n^d \to \mathbb{C}$, as the 
  discretization of a function $u$ on the $d$-dimensional torus of length size $nh$,
   $u:\bT^d_{nh} \to \bC$, with mesh size $h>0$, i.e., $\psi(x)=u(hx)$,
   for $x \in \bTd_n^d$. 
  Then the discrete  nonlinear Schr\"{o}dinger equation (DNLS) 
  is the following system of ODEs: 
  \begin{equation} \label{DNLSd}
    i\frac{d\psi(x,t)}{dt}=-\Delta_d \psi(x,t)+\kappa|\psi(x,t)|^{p-1}\psi(x,t),
    \qquad x \in \bTd_n^d
\end{equation}    
  where $\Delta_d$ is the $d$-dimensional discrete Laplacian: 
  $$
  \Delta_d \psi(x) =h^{-2} \sum_{|y-x|=1} \big(\psi(y)-\psi(x)\big).
  $$
  % where $|.|_d$
  % denotes the $L^{\infty}$ norm on $\bT_n^d$, and we imposed
  % periodic boundary condition $0 \equiv n$. 
  These equations conserve the energy, given by the Hamiltonian
  	\begin{equation} \label{Hamd}
  		{\mathcal H}_n(\psi)=   s\sum_{\substack{x,y \in \bTd_n^d, \\|x-y|=1}} 
  		\frac{h^{-2}}{2}|\psi(x)-\psi(y)|^2+
  		\frac{s\kappa}{p+1} \sum_{x \in \bTd_n^d}|\psi(x)|^{p+1},
              \end{equation}
              and the mass, given by the $\ell^2$ norm:
              	\begin{equation} \label{massd}
  		\cM_n(\psi)= s\sum_{x \in \bTd_n^d} |\psi(x)|^2.
  	\end{equation}
  	Here $s>0$ is a scaling parameter that we will choose opportunely later.
        % We will use the notation
        % ${\mathcal H}_n^1 = {\mathcal H}_n$ and $M_n^1 = M_n$.
        % (possibly depending on $h,n$, the natural 
  	% choice here is $s=h^d$, however, different 
  	% choices for $s,h$ leads to different large scale behaviors). 
  	% The other conserved quantity for the dynamics \eqref{DNLSd} is the mass 
  	% $M_n^s: \bC^{n^d} \to \bR$, \footnote{This quantity can be viewed as the number 
  	% of the particles in certain models.}, which is given by: 
  
  	% Observe that $M_n^s$ is conserved by the evolution \eqref{DNLSd}, 
  	% thanks to the periodic boundary condition.
  	 %$ d M_n(\psi)/dt=in[(\psi(n+1)\psi^*(n)-\psi^*(n+1)\psi(n))-(\psi(1)\psi^*(0)-\psi^*(1)\psi(0))]$ imposing the periodic boundary condition ($0\equiv n $ and $1 \equiv n+1$) we deduce the conservation of the mass. 
  	 Denote $\psi(x)=\psi_r(x)+i\psi_i(x)=|\psi(x)|e^{i\theta(x)}$, 
  	 the deterministic evolution equation \eqref{DNLSd} can be regarded 
  	 as a Hamiltonian dynamics with the following generator:
  	 \begin{equation} \label{generatorhamd}
           \cA=
           s^{-1} \sum_{x \in \bTd_n^d}(\partial_{\psi_i(x)} {\mathcal H_n})\partial_{\psi_r(x)}
           - (\partial_{\psi_r(x)} {\mathcal H_n}) \partial_{\psi_i(x)}.
	\end{equation}  	  
%	where $\mathcal{H}=\frac{1}{s}H_n$, and $H_n$ is defined in \eqref{Hamd}.  	
  	Moreover, define the operator $\partial_{\theta(x)}$ acting on a 
  	suitable function $F:\chi \to \mathbb{C}$ as 
  		\begin{equation} \label{phasederivative}
                  \partial_{\theta(x)}F(\psi)=
                  (\psi_r(x)\partial_{\psi_i(x)}-\psi_i(x)\partial_{\psi_r(x)})F(\psi).
		\end{equation}  	  
                Corresponding to a positive temperature $\beta^{-1}>0$,
                define:
 	\begin{equation} \label{generatorrandd}
          \cS=\beta^{-1}\sum_{x \in \bTd_n^d}
          e^{\beta {\mathcal H_n}}\partial_{\theta(x)}e^{-\beta {\mathcal H_n}}
          \partial_{\theta(x)}.
	\end{equation} 	  	
  	 
  	Fix  $\beta>0 , \gamma>0$, and consider the Markov process
        with values in $\chi$, generated by 
\begin{equation} \label{generatord}
  	L=\cA+ {\gamma} \cS, 
  \end{equation}
  where $\cS$ and $\cA$ are defined in
  \eqref{generatorrandd} and \eqref{generatorhamd}.
  Since $\partial_{\theta(x)}\psi(x)=i\psi(x)$, we have
  $$
  \cS \psi(x)=- \psi(x)\left(\beta^{-1} + i\partial_{\theta(x)} {\mathcal H}_n(\psi)\right)
  = - \psi(x)\left(\beta^{-1} + i s\ \text{Im} [\psi^*(x)\Delta_d \psi(x)]\right)
  $$  
  gives us the explicit form of system of
  stochastic differential equations generated by \eqref{generatord}:
	\begin{equation} \label{sde1d}
\begin{split}
  d\psi(x,t) =&i[\Delta_d \psi(x,t)-\kappa|\psi(x,t)|^{p-1}\psi(x,t)]dt
  -\gamma \psi(x,t)(\beta^{-1} + i\partial_{\theta(x)}{\mathcal H}_n(\psi))dt \\
&-i\sqrt{2 \gamma \beta^{-1}}\psi(x,t)dw(x,t),\qquad x \in \bTd_n^d,
\end{split}
	\end{equation}
        where $\{ w(x,t), x \in \bTd_n\}$ are real independent Wiener processes.
        
        We observed that $\cA \cM_n(\psi)=0$,
        one can check that $\cS \cM_n(\psi)=0$.
   Therefore, mass is a conserved quantity
   for the dynamics \eqref{generatord}. Hence, 
   if we assume the initial condition $\psi(0,t)=\psi_0 \in \bC^{n^d}$,  such that
   $\cM_n(\psi_0)=m$, then our dynamics is confined
   in the compact manifold with 
   $\cM_n(\psi)=m$, which is a $(2n^d-1)$-sphere.
   We denote this sphere by $S_{m,s}^n$:
  	\begin{equation} \label{configurationspaced}
          S_{m,s}^n=\{\psi \in \mathbb{C}^{n^d} | \cM_n(\psi)
          % =s\sum_{x \in \bTd_n^d} |\psi(x)|^2
                =m\}.
  	\end{equation}

        \begin{prop}\label{hypo}
          The generator $L$ is hypoelliptic.
        \end{prop}
        The proof follows from H\"{o}rmander characterization, i.e., that the Lie algebra 
        generated
        by $\{\cA, \partial_{\theta(x)}, x\in \bTd^d_n\}$ generates the tangent space
        of $S_{m,s}^n$. This is proven in Appendix \ref{app:hor}.

        % Since the configuration space $S_{m,s}^n$ is compact and coefficients are smooth,
        % one can deduce that the coefficients are 
	% Lipschitz and we have the global well-posedness. \\
	Let $d\mu_{m,s}^n$ be the uniform probability measure on $S_{m,s}^n$, 
	one can define this measure 
	as the projection of the Lebesgue measure on  $S_{m,s}^n$, properly normalized. 
	%  This measure can be considered as 
	% $\delta(M_n^s(\psi)-m)$ in distributional sense.
        Define the canonical Gibbs 
	measure with inverse temperature $\beta$ on $S_{m,s}^n$ as 
		\begin{equation} \label{gibbsmesured}
 			d\mu_{\beta,m,s}^n=\frac{1}{Z_n(\beta,m,s)} e^{-\beta {\mathcal H}_n(\psi)} 
 			d\mu_{m,s}^n,
		\end{equation}
  	% as a distribution we have: 
  	% \begin{equation} \label{Gibbs2d}
  	% 	d\mu_{\beta,m,s}^n=\frac{1}{Z_n(\beta,m,s)} e^{-\beta H_n(\psi)} \delta(M_n^s(\psi)-m)\prod_{x \in \bTd_n^d} d\psi_r(x)d\psi_i(x),
	% \end{equation}  	
	% where   $\Pi_{x \in \bTd_n^d} d\psi_r(x)d\psi_i(x)$ is the 
	% Lebesgue measure on $\mathbb{C}^{n^d}$. 
\\
	Here $Z_n(\beta,m,s)$ is the partition function:
  		\begin{equation} \label{partitionfunctiond}
			Z_n(\beta,m,s)= \int_{S_{m,s}^n} e^{-\beta {\mathcal H}_n(\psi)} d \mu_{m,s}^n.
		\end{equation}
	Note that, since $\mathcal H_n$ is a smooth function on a compact set and 
    therefore, bounded from below, $Z_n(\beta,m,s)$ is finite, and consequently, 
	the existence of $d\mu_{\beta,m,s}^n$ is evident.

	The observation that $\forall f \in C_b(S_{m,s}^n), 
	\int_{\mathbb{C}^n}Lf d\mu_{\beta,m,s}^n=0$, 
	implies that $d\mu_{\beta,m,s}^n$ is an invariant measure for the dynamics
	\eqref{generatord}(\eqref{sde1d}). In fact, if we fix $m,\gamma, \beta>0$ 
	this measure is 
	the unique invariant probability measure:
		\begin{theorem}
	Fix the parameters $h,s,\gamma>0$, 
	the mass of the field $m$, and inverse temperature $\beta>0$,  
	the measure $d\mu_{\beta,m,s}^n$ is the unique invariant 		
	measure	for the dynamics generated by \eqref{generatord}.
		\end{theorem} \label{thinmed}

                \begin{proof}
                  Without losing generality we can fix $h=s=1$. 
  Since the generator $L$ is \textbf{hypoelliptic}, the stationary measure must have density 
  w.r.t $d\mu^n_{m,s}$,
  and then also w.r.t $d\mu_{\beta,m,s}^n$. 
  Denoting $f(\psi)$ the density w.r.t  $d\mu_{\beta, m,s}^n$, 
  it must satisfy the equation
  \begin{equation}
    \label{eq:9}
    0 = L^* f = (-\cA + \gamma \cS) f,
  \end{equation}
	where $L^*$ denotes the adjoint of $L$  in $L^2(d\mu_{\beta,m,s}^n)$.    
	Since $L$ is hypoelliptic, $f$ is smooth and \eqref{eq:9}
    is valid pointwise.
    Multiplying by $f$ and integrating w.r.t $ d\mu_{\beta,m,s}^n$, we have
	\begin{equation}	\label{eq:10}
  		0 = {\gamma} < f (-\cS) f> = {\gamma} \sum_{x \in \bTd_n^d}
  		 <(\partial_{\theta(x)} f)^2>,
	\end{equation}
	where $<\cdot>$ denotes integration w.r.t $ d\mu_{\beta, m,s}^n$.
	This means that $\partial_{\theta(x)} f =0$ $d\mu_{\beta, m,s}^n$-a.e.,
	and $\cA f = 0$. We want to conclude that $f =1$, $d\mu_{\beta, m,s}^n$-a.e..
	Since $\partial_{\theta(x)} f =0$ for any $x$, then $f = \tilde
	f(|\psi(x)|^2, x \in \bTd_n^d)$. 
	The operator $\cA$ can be written as $\cA^0 + \cA^p$, with
	\begin{equation}
  	\label{eq:11}
  		\cA^0 = \sum_{x \in \bTd_n^d} \left\{\left(\Delta_d
      \psi_i(x)\right) \partial_{\psi_r(x)} - \left(\Delta_d
      \psi_r(x)\right) \partial_{\psi_i(x)}\right\},
	\end{equation}
	and
	\begin{equation}
  	\label{eq:12}
   	\cA^p = \kappa\sum_{x \in \bTd_n^d} |\psi(x)|^{p-1} \left\{
    \psi_i(x) \partial_{\psi_r(x)} - \psi_r(x) \partial_{\psi_i(x)}\right\} =
   	\kappa\sum_{x \in \bTd_n^d} |\psi(x)|^{p-1} \partial_{\theta(x)}.
	\end{equation}
	It is immediate that $\cA^p f =0$, hence, $\cA^0 f=0$, pointwise. 
	Let us denote $a(x):= |\psi(x)|^2$, and the canonical basis of $\bR^d$ by
	$\{e_j \}_{j=1}^d$, then we have: 
	\begin{equation}	\label{eq:13}
  	\begin{split}
  	&0= \cA^0 f =  2\sum_{x \in \bTd_n^d} \left\{\left(\Delta_d \psi_i(x)\right) \psi_r(x) -
    \left(\Delta_d \psi_r(x)\right) \psi_i(x)\right\}
   	\left[\partial_{a(x)}\tilde f \right] (|\psi(y)|^2, y \in \bTd_n^d) \\
  	&= 2\sum_{x \in \bTd_n^d} \sum_{j=1}^d
  	\nabla_+^j(\psi_r(x) \psi_i(x-e_j) - \psi_i(x) \psi_r(x-e_j))
   	\left[\partial_{a(x)}\tilde f \right] (|\psi(y)|^2, y \in \bTd_n^d) \\
  	&=2\sum_{x \in \bTd_n^d} \sum_{j=1}^d
  	\left[\psi_r(x) \psi_i(x-e_j) - \psi_i(x) \psi_r(x-e_j)\right]
  	\left[(\partial_{a(x)} -  \partial_{a(x-e_j)}) \tilde f \right](|\psi(y)|^2, y \in \bTd_n^d)\\
  % = 2\sum_x \left[\cos(\theta_x) \sin(\theta_{x-1}) - \sin(\theta_x) \cos(\theta_{x-1})
  % \right] |\psi(x)|  |\psi(x-1)| \left[(\partial_{a(x)} -  \partial_{a(x-1)}) \tilde f 		\right]
  % (|\psi(1)|^2, \dots, |\psi(n)|^2)\\
  &= 2\sum_{x \in \bTd_n^d} \sum_{j=1}^d  \sin(\theta_{x-e_j} - \theta_x)
 |\psi(x)|  |\psi(x-e_j)
 |\left[(\partial_{a(x)} -  \partial_{a(x-e_j)}) \tilde f \right]
 (|\psi(y)|^2, y \in \bTd_n^d),
	\end{split}
	\end{equation}
	where $\nabla_+^j$ denotes the discrete gradient in the $e_j$ direction
	 $(\nabla_+^j g)(x)=g(x+e_j)-g(x)$.
	Since this relation is true pointwise for any $\psi \in S^n_{m,s}$, by  choosing 
	a proper $\psi$ (for example one can 
	 take  $\theta_y$ equal to zero, for $y \in \bTd_n^d$, except $\theta(x)$, and
	 take $|\psi(y)|=0$, for all $|y-x|_d=1$, except $x-e_j$),
	 we have that 
	 \begin{equation} \label{eq:14}
	 (\partial_{a(x)} -  \partial_{a(x-e_j)}) \tilde f (a(y), y \in \bTd_n^d) = 0,
	 \end{equation} 
	 pointwise for every $x \in \bTd_n^d$,
         and any $1 \leq j  \leq d $.

         From \eqref{eq:14},
	 we conclude that 
	$(\partial_{a(x)} - \partial_{a(z)} ) \tilde f (a(y), y \in \bTd_n^d) =
	0$ for any $x,z \in \bTd_n^d$. This implies
        $$
        \tilde f (a(y), y \in \bTd_n^d)= 
	F \left(\sum_{y\in \bTd_n^d} a(y)\right) = F(m).
          $$
          which yields the result.
\end{proof}
	\begin{rem}
		Notice that the proof of Theorem \ref{thinmed}, works for any other 
		non-linearity of the form $F(|\psi|)$ with smooth $F$ (at least $C^2$). 	
	\end{rem}

        By classical theorems in control theory,
        given the H\"{o}rmander condition,
        and the existence of a unique invariant measure 
        with full support on $S^n_{m,s}$,
        it follows the strict positivity of the probability transition
        (cf. \cite{hairer2005probabilistic}, proof of Theorem 2.1)
        and the following proposition:
        
	 \begin{prop} \label{thlongtimed}
           Consider the dynamics
           which is generated by \eqref{generatord},
           denote the law of this
 process by $\mu_t^{\beta,n,m}$ with initial condition $\mu_0^{\beta,n,m}=\delta_{\psi_0}$, 
 where $\psi_0$ is an arbitrary element of $S_{m,s}^n$. There exist $C(n,m,\psi_0)$ 
 and $\gamma_0$, such that 
\begin{equation}
\|\mu_t^{\beta,n,m}-d\mu_{\beta,m,s}^n\|_{TV}\leq Ce^{-\gamma_0t}.
\end{equation} 
In particular, we have the weak convergence: 
\begin{equation}\label{Longtimebehavior1d}
\mu_t^{\beta,n,m} \mathop{\longrightarrow}_{t\to \infty} \mu_{\beta,m,s}^n.
\end{equation}
\end{prop}
\begin{proof}
Since $\mu_{\beta,m,s}^n$ is the unique invariant measure (ergodicity), with full support (for any open set 
$A \subset S^n_{m,s}$, 
$\mu_{\beta,m}^n(A)>0$), given the H\"{o}rmander condition we can use the 
result of \cite{hairer2005probabilistic}, (proof of Theorem 2.1 in  
\cite{hairer2005probabilistic}) and deduce the strict positivity of the probability transition.
 Furthermore, having the strict positivity of the probability transition, compactness
 of the phase space, as well as the
 hypoellipticity of the generator, we can conclude by Theorem 8.9 of \cite{bellet}.
\end{proof}
  The novelty of the stochastic perturbation  \eqref{generatorrandd} can 
  be described as follows:
  it's a  
  mass-conserving white noise, such that the Gibbs measure is the unique invariant 
  measure for the dynamics, 
  and it provides good ergodic properties as in Theorem \ref{Longtimebehavior1d}.
  This perturbation is quite "powerful" in the sense that its ergodic properties 
  do not depend on the non-linearity,
  % dimension and the parameters $s,h,\gamma$, 
  % and the fact that
  and we can consider either focusing or defocusing non-linearity. 
  In either of these cases the long time behavior is given
  by the corresponding Gibbs measure. 
  However, depending on the choice of parameters
  $d,s,h,\kappa$ many interesting 
  phenomena can be observed in the large scale limit. In the rest of this note, 
  we focus on one particular case:
  one-dimensional focusing nonlinear Schr\"{o}dinger 
  Equation on the torus.
  % This means we take $h= \frac{1}{n}$, $s=\frac{1}{n}$, $\kappa=1$, 
  % $p=3$, and $d=1$. From now on we fix these parameters,
  % and we study the long time
  %  and large scale limit of the corresponding dynamics.   
  	
  \section{Large Scale Limit  and Main Result}
  \label{sec:mainresult}
  \subsection{Preliminaries about periodic cubic
    nonlinear Schr\"{o}dinger equation}
  
 In this section, we recall rather basic results about the focusing nonlinear 
 Schr\"{o}dinger equation (NLS) with periodic boundary conditions. 
 Consider the following nonlinear cubic Schr\"{o}dinger equation:   
 \begin{equation} \label{NLS}
	\begin{split}     
     &i\partial_t \psi(x,t) =-\partial_{xx} \psi(x,t)-|\psi(x,t)|^2\psi(x,t), \: (t,x) \in \mathbb{R}_+\times \mathbb{R}, \\
	&\psi(x,0)=\psi_0(x),\: \psi_0 \in H^1(\bT_L),
	\end{split}
 \end{equation}    
 where we assume the periodic boundary conditions by the definition of $H^1(\bT_L)$
 as:
 \\ $H^1(\bT_L) = \{ u \in H^1_{loc}(\mathbb{R},\mathbb{C})|\: \forall x \in \mathbb{R},
 \quad u(x+L)=u(x)\}$, with the following norms and inner product 
 ($\bar{v}$ indicates the complex conjugate): 
 \begin{equation} \label{norm}
	\|u\|_{L^p}= \left(\int_{\bT_L}|u|^p dx\right)^{\frac{1}{p}}, \: \: (u,v)=\int_{\bT_L} u\bar{v} dx, \: 
	\: \: \|u\|_{H^1}= \left(\int _{\bT_L} \left( |\partial_xu|^2+|u|^2\right) dx \right)^{\frac12}.
 \end{equation}		  
 	Global wellposedness of this problem is established in \cite{bourgain}, 
 	\cite{cazenave03};
 in particular,  $\forall t>0, \: \psi(x,t) \in H^1({\bT_L})$. Note that this equation
 has two important conserved quantities \footnote{In fact, since this equation is
 completely integrable, we have infinite conserved quantities. However, most of the
 results in this note can be generalized to the sub-critical non-linearities 
 that are not integrable,
 i.e., we can change the  nonlinearity term in \eqref{NLS} into $|\psi|^{p-1}\psi$ with
 $1\leq p <5$. Notice that if $p \neq 3$, w do not have the explicit characterization 
 of the Solitions}: the energy or Hamiltonian $\cH$, and $L^2$ norm or mass $\cM$, defined by
	\begin{equation} \label{conserved quantities}
		\begin{split}
                  \mathcal{H}(\psi)=
                  \frac12 \int_{\bT_L} |\partial_x\psi|^2  dx- \frac14 \int_{\bT_L} |\psi|^4 dx, \qquad
		 	\mathcal{M}(\psi)= \int_{\bT_L} |\psi|^2 dx.
		 \end{split}
               \end{equation}
               
  One of the main features of this equation % , which we are interested in,
  is the
  existence of a special class of solutions called the \textit{"standing waves"} or
 \textit{"periodic waves"}. These are \textit{time periodic} solutions having the
  following form:
	\begin{equation} \label{Travelling wave1}
		\psi(x,t)=e^{i\omega t}u(x).
	\end{equation}
  If $\psi(x,t)=e^{i\omega t}u(x)$ be a solution of \eqref{NLS}, then $u(x)$ 
  should satisfy the following ODE, with periodic boundary condition:
	\begin{equation} \label{ODE1}
	  	u''(x)-\omega u(x)+|u(x)|^2u(x)=0.
              \end{equation}
               %\emph{Why we consider only real valued solutions?}
		Notice that the solution of \eqref{ODE1} characterizes the minimum of the energy  $
		\mathcal{H}(u)$, under the constrain $\cM(u) = m$, where 
		the frequency $\omega$ plays the role of Lagrange multiplier.
		
                In general, we should consider complex valued solutions of
                \eqref{Travelling wave1}. On the other hand, writing this solution as
                $u(x) = \rho(x) e^{i\theta(x)}$, the corresponding energy is given by
                $$
                \mathcal{H}(u)=
                \frac12 \int_{\bT_L} \left( |\rho'(x)|^2 + \rho(x)^2 |\theta'(x)|^2\right)  dx
                - \frac14 \int_{\bT_L} |\rho(x)|^4 dx.
                $$
                This shows that the minimum of the energy  $\mathcal{H}(u)$, under
                the constrain $\cM(u) = m$ is attained for $\theta(x) = \text{constant}$.
                %The frequency $\omega$ plays the role of Lagrange multiplier, and the
                %solution of \eqref{ODE1} characterize such minimum.
                Consequently,
                this minimum are defined up to a constant phase and we can
                choose positive real solutions.
                Also notice that translations $u_y(x) = u(x+y)$ do not change 
                energy and mass.

	% One can find different class of smooth solutions for this equation
	% (cf. \cite{GLT}, \cite{pava}, \cite{gallay}) using the properties
        % of Jacobi elliptic functions ($\cn(\cdot) ,\sn(\cdot), \dn(\cdot)$)
	% In the rest of this note, we do not use the specific form of these functions, so we
	% introduce them in Appendix(??).
        Here, if we fix the $L$, and assume
	 $u$ to be real-valued, and positive, and fix the mass of $u$ to be $\cM(u)=m$, 
	 then under these assumptions, \eqref{ODE1} has a unique (up to a translation) 
	 smooth solution, this solution can be written in terms of Jacobi elliptic functions
	 as  $u(x)=\alpha \dn(\lambda x,k)$, where $k \in (0,1), \alpha$, and $\lambda>0$,
	 $\omega>0$ are
	 uniquely determined by $m$, and $L$ (cf. \cite{GLT}, \cite{pava}, \cite{gallay}, 
	 cf. Appendix \ref{JEF} for the definition of $\dn$).
       %  {\color{red} \emph{($\omega$ is determined by $m$ and $L$!)}}.
         We recall the following crucial result from \cite{GLT}, Proposition 3.2,
         which characterizes this solution as the minimizer of
	 $\cH(\psi)$ under the constraint that $\cM(\psi)=m$. 
 		\begin{theorem} \label{thMinimize}
 		Fix $m,L \in \mathbb{R}_+$, and consider the following minimization problem: 
 			\begin{equation} \label{Minimiztion}
  			E_0(m,L):=\inf \{\mathcal{H}(u) | \mathcal{M}(u)=m, \: u \in H^1({\bT_L})\},
 			\end{equation}
		then we  have: $- \infty< E_0(m,L)<0$, and
		\begin{enumerate}
			\item If $0<m\leq\frac{\pi^2}{L}$, then the constant function
			 $Q_{m,L}(x)=(\frac{m}{L})^\frac12$ is the unique minimizer of
			 \eqref{Minimiztion}.
                         This uniqueness is up to a multiplication by a constant phase.
                       \item If $\frac{\pi^2}{L} < m$, then
                         $Q_{m,L}(x):=\alpha \dn(\lambda x,k)$ is
			 the unique minimizer of \eqref{Minimiztion}, up to a translation and
                         multiplication by a constant phase. Moreover,
                         $\alpha,\lambda>0 , k \in(0,1)$ are determined uniquely by $m,L$. 
		\end{enumerate}
                Furthermore, we have compactness of the minimizing sequence
                up to a phase shift and
	 translation in $H^1({\bT_L})$, i.e., for any sequence $u_n$ in 
	 $H^1({\bT_L})$, such that $\mathcal{H}(u_n) \to E_0(m,L)$, as $n \to \infty$,
	 there is a subsequence $u_{n_k}$, and sequences $\gamma_k \in [0,2\pi)$,
         and $x_k\in {\bT_L}$,  where $e^{i\gamma_k}u_{n_k}(.+x_k) \to Q_{m,L}$, in $H^1({\bT_L})$. 

\end{theorem}  
	Since each solution of \eqref{ODE1} (and consequently a solution to
	\eqref{NLS}) corresponds to the minimization problem \eqref{Minimiztion},
	 by abusing the terminology, we use the term "\textit{standing wave}" 
	or \textit{Soliton} for $Q_{m,L}$. % Notice that the terms \emph{Soliton} refers to the
	% counterpart of this function on whole space. However, we use the same term for 
	% $Q_{m,L}$.

        { Notice that multiplying \eqref{ODE1} by $\bar u$ and integrating,
          we obtain the following relation
          \begin{equation*}
            	E_0(m,L) = \frac 14 \int_{\bT_L} u^4(x) dx - \frac{\omega m}{2}.
              \end{equation*}
              that implies $\omega \ge \frac 1{2m} \int_{\bT_L} u^4(x) dx + \frac{m}{2L^2}$.
}

\subsection{Stochastic perturbation of discrete focusing NLS}

	In this section, we are going to perturb the NLS \eqref{NLS}, with the stochastic 
	heat bath, which we defined in Section \ref{SDsec},
        namely \eqref{generatorrandd}.
        Without loosing generality, in order to simplify notation, we fix the
        macroscopic length $L=1$.
	This means that
        we fix the following parameters $h=\frac{1}{n}, s=\frac{1}{n}, d=1, p=3, \kappa= -1$.
        % This corresponds to fix the macroscopic length $L=1$.
	% As before, we spatially discretize in order to avoid technicality concerning
	% well-posedness of the dynamics. 
        Here, we briefly recall the dynamics of Section
	\ref{SDsec} in this particular setup, in order to set the notations.  	 
  	\\
   Fix $n \in \mathbb{N}$, the configuration space is $\chi = \mathbb{C}^n$  
  and denote a typical element of $\chi$ by $\{ \psi(x) \}_{x \in \bTd_n}$, 
  with $\bTd_n=\{1,2,\dots,n\}$ is the discrete torus of size $n$. 
  Equivalently, a function $\psi$ on $\bTd_n$ can be seen as
  discretization of a function $u$ on a unit torus,
   $u:\bT \to \bC$, with mesh size $\frac{1}{n}$, i.e., $\psi(x)=u(\frac{x}{n})$,
   for $x \in \bTd_n$.
  Then the discrete cubic focusing nonlinear Schr\"{o}dinger equation (DNLS) 
  is the following system of ODEs: 
  \begin{equation} \label{DNLS}
  i\frac{d\psi(x,t)}{dt}=-\Delta \psi(x,t)-|\psi(x,t)|^2\psi(x,t),
\end{equation}    
  where $\Delta \psi(x,t) =n^2 (\psi(x+1)-2\psi(x)+\psi(x-1))$, and we imposed
  periodic boundary condition $\psi(0) \equiv \psi(n)$. 
  Notice that we define $\Delta$ such that formally in the limit $n \to \infty$, 
  this definition coincides with the continuous Laplacian on a unit torus. 
  
  Similar to the continuous case, we have the energy or
  Hamiltonian $\mathcal H_n:\bC^n \to \bR$ as a conserved quantity,
  that is defined by:
  	\begin{equation} \label{Ham1}
  	\mathcal 	H_n(\psi)= \frac1n \sum_{x \in \bTd_n} \frac{n^2}{2}|\psi(x)-\psi(x-1)|^2-
  		\frac1{4n} \sum_{x \in \bTd_n}|\psi(x)|^4 = G_n(\psi) - V_n(\psi) ,
  	\end{equation}
  where we have denoted the
  	kinetic energy $G_n(\psi)$ and the potential energy $V_n(\psi)$ as:
  	\begin{equation} 
  	\begin{split}
		G_n(\psi)= \frac1n \sum_{x \in \bTd_n} \frac{n^2}{2}|\psi(x)-\psi(x-1)|
		^2, \label{grad1} \qquad
		V_n(\psi)=\frac{1}{4n} \sum_{x \in \bTd_n}|\psi(x)|^4. 
  	\end{split}
	\end{equation}  	 
  	The other conserved quantity is given by the mass 
  	$\cM_n: \bC^n \to \bR$, defined by: 
  	\begin{equation} \label{mass}
  		\cM_n(\psi)=\frac1n\sum_{x \in \bTd_n} |\psi(x)|^2.
  	\end{equation}

  	 Notice that we scaled \eqref{Ham1} and \eqref{mass}, such that in the limit
  	 as $n \to \infty$, we recover $\cH$, and $\cM$ formally. %$ d M_n(\psi)/dt=in[(\psi(n+1)\psi^*(n)-\psi^*(n+1)\psi(n))-(\psi(1)\psi^*(0)-\psi^*(1)\psi(0))]$ imposing the periodic boundary condition ($0\equiv n $ and $1 \equiv n+1$) we deduce the conservation of the mass. 

         The stochastic perturbation we consider will only conserve the mass.
  	 Recall $\psi(x)=\psi_r(x)+i\psi_i(x)=|\psi(x)|e^{i\theta(x)}$, 
  	 the generators of the Hamiltonian and stochastic noise at temperature $\beta^{-1}$
  	  read 
  	 \begin{equation} \label{generatorham}
           A_n=  n\sum_{x \in \bTd_n}(\partial_{\psi_i(x)} \mathcal{H}_n)\partial_{\psi_r(x)}
           - (\partial_{\psi_r(x)} \mathcal{H}_n) \partial_{\psi_i(x)},
	\end{equation}  	  
	% where $\mathcal{H}_n =nH_n$, and $H_n$ is defined in \eqref{Ham1}.  	
  	\begin{equation} \label{generatorrand}
 		\cS_n=  \beta^{-1}\sum_{x \in \bTd_n} e^{\beta \cH_n}\partial_{\theta(x)}e^{-
 		\beta \cH_n} \partial_{\theta(x)}.
	\end{equation} 	  	
  	 
  	Fix  $\beta>0 , \gamma>0$, then the generator of the dynamics and corresponding 
  	 system of 
  	stochastic partial differential equations with values in $\chi$, are as follows:
\begin{equation} \label{generator}
  	L_n=A_n+ {\gamma} \cS_n, 
  \end{equation}
	 
	\begin{equation} \label{sde1}
\begin{split}
  d\psi(x,t) =&i[\Delta \psi(x,t)+|\psi(x,t)|^2\psi(x,t)]dt -\gamma \psi(x,t)(\beta^{-1}
  +i\partial_{\theta(x)} \cH_n(\psi))dt \\
&-i\sqrt{2\gamma \beta^{-1}}\psi(x,t)dw(x,t),
\end{split}
	\end{equation}
  where $\{ w(x,t), x \in \bTd_n\}$ are real independent Wiener processes. 
  
  Due to the mass conservation, having an 
  initial condition $\psi(0,t)=\psi_0 \in \bC^n$  such that
  $\cM_n(\psi_0)=m$, our dynamic will be confined in the sphere
   $S_m^n=\{\psi \in \mathbb{C}^n | \cM_n(\psi)=m\}$.
	Denote the uniform probability measure on $S_m^n$ by $d\mu_m^n$, and 
	 define the canonical Gibbs 
	measure with inverse temperature $\beta$ on $S_m^n$ as 
		\begin{equation} \label{gibbsmesure}
 			d\mu_{\beta,m}^n=\frac{1}{Z_n(\beta,m)} e^{-\beta \cH_n(\psi)} d\mu_m^n,
		\end{equation}
  	Here $Z_n(\beta,m)= \int_{S_m^n} e^{-\beta \cH_n(\psi)} d \mu_m^n$.
	As we observed,  $Z_n(\beta,m)$ is finite, and consequently, the existence of
	$d\mu_{\beta,m}^n$ is evident, since $\cH_n(\psi)$ is bounded from below in 
    $S_m^n$. 
      However, one can find a lower bound for $\cH_n(\psi)$, which is uniform in $n$, 
     using a version of Gagliardo-Nirenberg inequality in the discrete periodic setup.
     This will be discussed broadly in the Section \ref{Dissol} and Appendix \ref{secGN}. 

	Applying the result of Section \ref{SDsec}, we have the following results:
	 By Theorem \ref{thinmed}
	 we know that $d\mu_{\beta,m}^n$ is the unique invariant measure for the dynamics
         \eqref{generator}(\eqref{sde1}).
	 Moreover, Proposition \eqref{thlongtimed} states that if $\mu_t$ 
	   denotes the law of the
 	process at time $t\ge 0$, generated by \eqref{generator}, with initial
 	 condition $\mu_0=\delta_{\psi_0}$, 
         where $\psi_0$ is an arbitrary element of $S_m^n$,
         then there exist $C(n,m,\psi_0)$ and $\gamma_0$ such that 
	\begin{equation}
	\|\mu_t - \mu_{\beta,m}^n\|_{TV}\leq Ce^{-\gamma_0 t}.
	\end{equation} 
	% In particular, we have
	%  %we have the weak convergence: 

	%  \begin{equation}\label{Longtimebehavior1}
	%   \mu_t^{\beta,n,m} \mathop{\longrightarrow}_{t\to\infty} d\mu_{\beta,m}^n.
	%  \end{equation}

  % Finally, we are prepared to state the main result of this manuscript. 
 If we run our dynamics for a long time, then 
   take the limit of large $n$ and small temperature $\beta^{-1}$ properly, 
   we end-up near \textit{Solitons} or \textit{standing waves} 
   ($Q_{1,m}$ from Theorem \ref{thMinimize}), with probability one.
   Notice that here we can take the limit in $\beta$ and $n$ simultaneously, where 
   we scale $\beta$ by a factor of $\vartheta(n)$. 
   In order to make these words rigorous, 
    and connect the discrete setup to the continuous one, 
   we need to introduce some notations.
  	For any   $\psi_n \in \bC^n$, we define its linear interpolation 
  	$\bar{\psi}^n: \bT \to \bC$, on a unit torus by % (Since in this section 
  	% we take the limit in $n$, we add the subscript $n$ for $\psi_n$ and $\bar{\psi}_n $):
  	\begin{equation} \label{linearinterpolation1}
	\begin{split}  	 	
  	 	\bar{\psi}_n(y)= &\psi_n \big([ny]\big)\big([ny]+1-ny\big)+\psi_n\big([ny]+1\big)
  	 	\big(ny-[ny]\big)
  	 	% \\&=\psi_n\big([ny]\big)+\big(ny-[ny]\big)\Big(\psi_n([ny]+1\big)-\psi_n\big([ny]
  	 	% \big)\Big)
                , \quad  \forall y \in \bT,
		\end{split}	
	\end{equation}  	
  	where $[ny]$ denotes the greatest integer less than $ny$.
  	Denote $H^1(\mathbb{T})$ by $H^1_{per}([0,1])=H^1_{per}$.
        For $x \in \mathbb{T}$, let $\tau_x$
	denotes the translation operator on $H^1_{per}$, i.e. $(\tau_xf)(y)=f(x+y)$,
        then, in order to deal with the phase multiplication and translation,
        define the following seminorm as in \cite{CH}:
	\begin{equation} \label{seminorm}
		\forall f,g \in H^1_{per}, \quad \|f-g\|_{\tilde{H}^1_{per}}:=\inf_{\gamma \in 
		[0,2\pi],x \in \mathbb{T}} \|e^{i\gamma}\tau_xf-g\|_{H^1_{per}}.
	\end{equation}
	
        In the following we set $Q_{1,m}=:Q_m$.
	Now we can state the main theorem of this section: 

\begin{theorem} \label{thlimit}
  Fix $m>0$, $\gamma>0$, and $\beta>0$, let $\beta_n=\vartheta(n) \beta $, 
  where $\vartheta(n)>0$ is a scaling parameter, such that 
  \begin{equation} \label{scaling}
  \lim_{n \to \infty} \frac{\vartheta(n)}{n} \to \infty.
  \end{equation}
  Let $\mu_t^{\beta_n,n,m}$ be the law of the process
  given by its generator \eqref{generator},   
  with the initial condition $\mu_0^{n,m}=\delta_{\psi_0^{n,m}}$, where $\psi_0^{n,m}$ 
  is a sequence of proper initial conditions, i.e., for all $n$, $\psi_0^{n,m} \in S^n_m$.
  %  Let $L=1$, and recall the definition of $Q_{1,m}=:Q_m$ 
  % from Theorem \ref{thMinimize},
  Then $\forall \: \epsilon>0$, we have:
\begin{equation} \label{limitmain1}
  \lim_{n \to \infty} \lim_{t \to \infty} \mu_t^{\beta_n,n,m}
  \Big(\|\bar{\psi}_n-Q_m\|_{\tilde{H}^1_{per}} <\epsilon \Big) \to 1.
\end{equation}
\end{theorem}  	
We briefly sketch the proof:
we have already proved that $\mu^{\beta_n}_{n,m}$ is the limit in
$t$ of  $\mu_t^{\beta_n,n,m}$.
Consequently, all we have to prove is that
\begin{equation}
  \label{eq:1}
  \lim_{n \to \infty}\mu^{\beta_n}_{n,m}\Big(\|\bar{\psi}_n-Q_m\|_{\tilde{H}^1_{per}}
  <\epsilon \Big) \to 1.
\end{equation}
% n \eqref{limitmain1}  converges to $\mu^{\beta_n}_{n,m}(|\bar{\psi}_n-Q_m|
 %         _{\tilde{H}^1_{per}} <\epsilon )$, as a result of Theorem \ref{thlongtimed},
 %         in particular  \eqref{Longtimebehavior1}. 
	 We can prove that the measure 
	  $\mu^{\beta_n}_{n,m}$ concentrates all its mass on the (discrete) configurations 
	  having close to the minimal energy,
          when we send temperature to zero with a proper speed.
	  It turns out that the proper speed here is to scale 
	  $\beta$ by $\vartheta(n)$, satisfying \eqref{scaling}.
	   Finally, we show that if a configuration has energy close to 
	  the minimal, it will be close to $Q_m$ in the sense of \eqref{limitmain1}. 
	  This can be done by adapting certain form of concentration compactness argument 
	  to the discrete setup.

          \begin{rem}\label{rem:drift}
            \textbf{About the exchange of limits in \eqref{limitmain1}:}
            In the evolution equation \eqref{sde1} the drift term
            $\partial_{\theta(x)} \cH_n(\psi) = \frac 1n \mathcal Im[\psi(x)\Delta\psi^*(x)]$
            would became very singular when $n\to\infty$ keeping the temperature
            positive. But with $\beta_n \to \infty$
            fast enough the solution should became enough regular in space so that the
            corresponding limit as $n\to\infty$ should be given by the continuous deterministic
            NLS. This will be investigated in a future work \cite{amir-so2}.
            {The later suggests that one could study the joint limit 
            $n,t \to \infty$, with $t_n=n^{\alpha}t$. We address the case $t_n\ll \beta_n$
            in \cite{amir-so2}. However, the case $t_n \gg\beta_n$ seems more challenging.}
             % If we start with an initial configuration typical for the canonical measure
            % $\mu^{\beta_n}_{n,m}$, then this will converge to $Q_m$ as $n\to\infty$,
            % and the limit $t\to\infty$ will be trivial.
            
          \end{rem}

  	\section{Discrete "Soliton"} \label{Dissol}

	As we already observed in Theorem \ref{thMinimize}, the function $Q_{m}$
	(Solitons) can be characterized as the minimizer of a certain variational problem,
	where we have the compactness of the minimizing sequence.
        Therefore, one can observe
	that for a function $u \in H^1_{per}([0,1])$, with $\cM(u)=m$, having "close to
	minimal" energies, means the function itself is close to $Q_{m}$ in the following
	sense:  
	\begin{lemma} \label{lemeenergycompactness}
		Assume $u \in H^1_{per}$ and $\mathcal{M}(u)=m$, then, $\forall 
		\epsilon>0, \: \exists \: \delta>0$, such that if $\mathcal{H}(u) \leq E_0(m) + 
		\delta$, then there exists $\gamma \in [0,2\pi]$, $x \in [0,1]$, such that
                $\|e^{i\gamma}u(.+x)-Q_{m}\|_{H^1_{per}} <\epsilon$, equivalently
                $\|u(x)-Q_{m}\|_{\tilde{H}^1_{per}} <\epsilon.$
	\end{lemma} 
	\begin{proof}
	This is straightforward by the compactness of the minimizing sequence in Theorem
	 \ref{thMinimize}.%  \textcolor{red}{This can be omitted}. We proof by contradiction: 
	 % If this is not true, then $\exists \: \epsilon>0$ such that
	 % $\forall \: \delta>0,$ there exist $u_{\delta} \in H_{per}^1$ 
	 % with $\mathcal{M}(u_{\delta})=m$, and $\mathcal{H}(u_{\delta}) \leq E_0(m) + \delta$ 
	 % such that   $ \forall \gamma \in [0,2\pi]$, $x \in [0,1]$,  $|e^{i\gamma}u(.+x)-		
	 % Q_{1,m}| _{H^1_{per}} >\epsilon$, now for $k \in \mathbb{N}$, take $\delta=\frac{1}
	 % {k}$, and construct a sequence $u_k$, so $\mathcal{H}(u_k) \to E_0(m)$. As a result of
	 % Theorem \ref{thMinimize}, there exist a subsequence $u_{k_n}$ and sequences
	 % $\gamma_n,x_n$, such that $e^{i\gamma_n}u_{k_n}(.+x_n) \to Q_{m,1}$ in $H^1_{per}$,
	 % but this is in contradiction with the assumption that 
	 % $\forall \gamma \in [0,2\pi]$, $x \in [0,1]$  $|e^{i\gamma}u(.+x)-Q_{1,m}|
	 % _{H^1_{per}} >\epsilon$.  and this proves the lemma.
	\end{proof}

 	 Similar to \eqref{Minimiztion}, fix $n>1 ,m>0$ and define $E_0^n(m)$  as
 	 follows: 
 		\begin{equation} \label{discreteminimization}
 			E_0^n(m) := \inf \{ \cH_n(\psi_n)|\psi_n \in S_m^n  \}.
 		\end{equation}
               % where $\cH_n(\psi), \cM_n(\psi)$ are defined in \eqref{Ham1}, \eqref{mass}.

                % {\color{red}
                %   \begin{lemma}
                %  $E_0^n(m) \ge E_0(m)$ and 
                %   $E_0^n(m) \mathop{\downarrow}_{n\to\infty} E_0(m)$
                % \end{lemma}
                % \begin{proof}
                %   For each $u\in H^1_{per}$
                % \end{proof}
                %   }
                
	Since $\cH_n(\psi_n)$ is a continuous function from the compact set 
	$S_m^n$ to $\mathbb{R}$, the image of this function is compact, hence, 
	$-\infty<E_0^n(m)$, and this infimum is achieved in a compact set,
	which will be called the set of "discrete Solitons" and denoted by 
	$\emptyset \neq \mathcal{Q}^n_m \subset S^n_m$.
        By the same argument as in the continuous case, discrete solitons
        are real-valued and positive up to a constant phase.
   	
	For $\psi_n \in S_m^n$, we define
        $\|\psi_n\|_{\ell^p(\bTd_n)}^p = \frac 1n \sum_j |\psi_n(j)|^p$.
        Then we can write
        $\cH_n(\psi_n)=G_n(\psi_n)-\frac{1}{4}\|\psi_n\|_{\ell^4(\bTd_n)}^4$,
        and by using the discrete Gagliardo-Nirenberg inequality 
	\eqref{GNdiscreteperiodicm}, we have: 
	\begin{equation} \label{energylowerbound}
			-\theta(m) \leq E_0^n(m) <0,
	\end{equation}		
	where % $\theta(m)$ is a constant depending on $m$ and $C$ (the constant appearing in 
	% \eqref{GNdiscreteperiodicm}), and independent of $n$. We can actually have  
	% 	the following explicit expression
                $\theta(m)=\frac{C^2}{64}m^3+\frac{C}4 m^2$. 
		First inequality is  a direct consequence of \eqref{GNdiscreteperiodicm}, and the 
		second one can be deduced	by considering the constant function $\psi_n(x)=
		\sqrt{m}$, for 	all $x \in \bTd_n$. 
%	\begin{equation} 
%\frac14 |\psi_n|_{L^4(\mathbb{T}_n)}^4 \leq \frac{c}{4}(m^{\frac32}G_n(\psi_n)^{\frac12}%+m^2) \implies
%H_n(\psi_n) \geq G_n(\psi_n)- \frac{c}{4}m^{\frac32}G_n(\psi_n)^{\frac12}-\frac{c}{4}m^2

%\end{equation}

%Since $G_n(\psi_n) \geq 0$ take $x=G_n(\psi_n)^{\frac12}$. The expression on the RHS of \eqref{energylowerbound} is equal to $x^2-c'm^{\frac{3}{2}}-c'm^2$, $x>0$ the minimum for this expression is a lower bound for $H_n(\psi_n)$ where will be achieved in $x=\frac{c'}{2}m^{\frac32}$ and is equal to $\theta(m)=-\frac{c'}{4}m^3-c'm^2$. So we obtain a lower bound $\theta(m)$ for $E^n_0(m)$ uniform in $n$. Moreover the simple test function $\psi_n(x)=\sqrt{m}$ implies $E_0^n(m)<0$.\\
	From \eqref{energylowerbound} we establish a simple but useful lemma:
  \begin{lemma}\label{boundongrad}
   		For every $\epsilon >0$, there exists $C(m,\epsilon)$,
   		such that for all $n \in \bN$ and  $\psi_n \in S^m_n$, with 
                $\cH_n(\psi_n) \leq E_0^n(m)+ \epsilon$,
                we have $G_n(\psi_n) \leq C(m,\epsilon)$.  
   \end{lemma}
	\begin{proof}
 	Consider the inequality \eqref{energylowerbound}, and \eqref{GNdiscreteperiodicm};
 	 denote $x=G_n(\psi_n)^{\frac12}$ so 	
 	$x \geq 0$. If $\cH_n(\psi_n) \leq E_0^n +\epsilon$ then, thanks to
 	\eqref{GNdiscreteperiodicm}  we  have
 	\begin{equation}
 		E_0^n(m) +\epsilon \geq x^2-c'm^{\frac32}-c'm^2 \implies x \leq
 		 C^{\frac12}(m, \epsilon). 
 	\end{equation}
	where $C^{\frac12}(m, \epsilon)$ is given by 
	$C^{\frac12}(m, \epsilon)=\frac{c'm^3+\sqrt{c'm^3+4(c'm^2+E_0^n(m)+\epsilon)}}{2}$,
        where the expression 
	under the square root is clearly positive, thanks to the expression of $\theta(m)$. 
	\end{proof}

        % When $n$ is large the uniform measure on 
	% $S^m_n$ concentrate on configurations with energy of $O(n)$
         Lemma \ref{boundongrad} states that if the energy is "small" ($O(1)$), then the
	 configuration should be "smooth" i.e., $G_n \sim O(1)$.
	 
	In the rest of this section, we prove % a result similar to Lemma 
	% \ref{lemeenergycompactness} for discrete configurations
	% with minimal or close to minimal energies. In fact, we prove
        that $\bar\psi_n$, the linear 
	interpolation of a configuration $\psi_n$, is arbitrarily close to $Q_{m}$ in 
	$\tilde{H}^1_{per}$, if we take $n$ sufficiently large, and the energy of $\psi_n$, 
	$\cH_n(\psi_n)$, sufficiently close to $E_0^n(m)$. The proof relies on the fact
	that the configurations with close to minimal energies are smooth in the sense that 
	their linear interpolation's norm ($L^p,H^1$ or even the energy) 
	is close to the corresponding discrete norms. This result heavily depends 
	on the inequality of Appendix \ref{secGN}. We begin by stating this result: 

		\begin{prop} \label{propdiscompact}
			Fix $m>0$, for any $\epsilon>0$, there exists $\eta(\epsilon)$ 
			and	$N_0(\epsilon)$, such that for $n>N_0(\epsilon)$, if 
			$\cH_n(\psi_n) \le E_0^n(m) + \eta$, then we have:
			$\|\bar{\psi}_n-Q_{m}\|_{\tilde{H}^1_{per}} < \epsilon$. 		
		\end{prop}

	 We divide the proof of \eqref{propdiscompact}, into a couple of simple lemmas.
         The advantage of the linear interpolation \eqref{linearinterpolation1} is that
         it conserves the kinetic energy, i.e.,
         $G_n(\psi_n) = \frac 12\int_0^1 |\partial_x\bar\psi_n|^2$.
         But unfortunately, in general, we have $\|\psi_n\|_{\ell^p(\bTd_n)}\ge 
         \|\bar\psi_n\|_{L^p}$
         for $p\ge 1$, thanks to the Jensen inequality. Consequently,
         in general we have $\cH_n(\psi_n) \le \mathcal{H}(\bar{\psi}_n)$
         and $\cM_n(\psi_n) \ge \mathcal M(\bar\psi_n)$.
         However, 
	the following lemma helps to establish the fact that these quantities are "close",
	for configurations with near minimal energies. 
	\begin{lemma} \label{lpdiccont}
	% Fix $m>0$, for any $c>0$ (independent of $n$) and $p>1$, recall the definition of 
	% $G_n$ \eqref{grad1}, then there exist $C_2(c,m,p)$ (independent of $n$) such that
	For all $n \in \bN$, if $\psi_n \in S_m^n$  % and $G_n(\psi_n) \leq c$,
        we have:
	$$
      \left|  \|\bar{\psi}_n\|^p_{L^p(\mathbb{T})}-\|\psi_n\|^p_{\ell^p(\bTd_n)} \right|\leq 
	\frac{p (2G_n(\psi_n))^{1/2}\left(m^{1/2} +G_n(\psi_n)^{1/2}\right)^{p-1}}{n},
        $$ 
	% where we recall the definition of $\bar{\psi_n}: \bT \to \bC$ as the linear 
	% interpolation of $\psi_n$ \eqref{linearinterpolation1}.	
	\end{lemma}
 	\begin{proof}
          We have $\psi_n \in S^m_n$, and define $\ell_n = \min \{ |\psi(x)| \: \big| x \in \bTd_n \}$,
          clearly 
 		$\ell_n \leq \sqrt{m}$. Moreover, for any $x \in \bTd_n$, we have:
 		$$|\ell_n-\psi_n(x)| \leq \sum_{j=1}^{n} |\psi_n(j)-\psi_n(j-1)| \leq
 		\sqrt n \left(\sum_{j=1}^{n} |\psi_n(j)-\psi_n(j-1)|^2\right)^{1/2} = \sqrt{ 2 G_n(\psi_n)},
               $$
               where we used a Cauchy Schwartz inequality.
               Therefore, we can deduce that  
               $$
               \sup_x |\psi(x)| \leq c_1=m^{\frac{1}{2}}+ (2G_n(\psi_n))^{\frac12}.
               $$
               Moreover, thanks to the definition of $\bar{\psi}_n(y)$,
               we have $|\bar{\psi}_n(y)| \leq c_1$, for all $y \in \bT$. 
 		Then we can simply compute: 
 		\begin{equation} \label{lp:1}
		\begin{split} 		
                  \left|	\|\bar{\psi}_n\|^p_{L^p(\mathbb{T})}-\|\psi_n\|^p_{\ell^p(\bTd_n)} \right|
                  \leq &
 		 \sum_{x=0}^{n-1}\int_{\frac{x}{n}}^{\frac{x+1}{n}}\Big||\psi_n(x)|^p-
 		 |\bar{\psi}_n(y)|^p\Big|dy\\
                 \leq &p c_1^{p-1} \sum_{x=1}^n\int_{\frac{x-1}{n}}	
 		 ^{\frac{x}{n}} \Big||\psi_n(x)|- |\bar{\psi}_n(y)|\Big|dy\\
 		 & \leq \frac{p c_1^{p-1}}{n} \sum_{x=1}^n |\psi_n(x)-\psi_n(x-1)| 
 		  \leq \frac{p c_1^{p-1} (2G_n(\psi_n))^{\frac12}} {n},  
		\end{split}		
		\end{equation} 		   
		where the first inequality comes from the  
		definition,  in the second inequality we used the fact that $\psi_n(x)$ and
		$\bar{\psi}_n(y)$ are bounded uniformly in $x$ and $y$, and in the
		third  inequality we used the definition of $\bar{\psi}_n(y)$: 
		$$
                \left| |\psi_n(x)|- |\bar{\psi}_n(y)|\right| \leq |\psi_n(x)-\bar{\psi}_n(y)|
		\leq |\psi_n(x)-\psi_n(x+1)|.
                $$
		Notice that the last inequality in \eqref{lp:1} is obtained as above.
	\end{proof}
 	As a straightforward consequence of Lemma \ref{lpdiccont}, we can deduce 
 	the following corollaries:

 	\begin{corollary} \label{energydisccont}
		% Fix $m>0$, recall the definition of $G_n$, $H_n$, $\cH$ and 
		% linear interpolation $\bar{\psi}_n$ \eqref{grad1}, \eqref{Ham1}, 
		% \eqref{conserved quantities}, and \eqref{linearinterpolation1}, 
		For any $c>0$, there exist $C_1(c,m)$, such that for every $n \in \bN$, 
		and $\psi_n \in S_m^n$, such that $G_n(\psi_n)<c$, then 
		$|\mathcal{H}(\bar{\psi}_n)-\cH_n(\psi_n)| \leq \frac{C_1(c,m)}{n}$.
	\end{corollary}
		\begin{proof}
		Thanks to the definition of $\bar{\psi}_n$ \eqref{linearinterpolation1}, the 
		weak derivative of $\bar{\psi}_n$ is given as follows: 
		 for any $y \in [0,1]$, if $\frac{x}{n}\leq y<\frac{x+1}{n}$ with 
		 $x \in \bTd_n$, then  
		 $\partial_y \bar{\psi}_n(y) = n(\psi_{n}(x+1)-\psi_n(x))$. Therefore, we have:
		 $\frac12 \int_0^1|\partial_y \bar{\psi}_n|^2= \frac{n}{2} \sum_{x=1}^n 
		 |\psi(x)-\psi(x-1)|^2=G_n(\psi_n)$. Hence, we have: 
		 $$
                 \left|\cH_n(\psi_n) -\cH(\bar{\psi_n})\right| =
		  \frac14 \Big|
		  \|\bar{\psi}_n\|^4_{L^4(\mathbb{T})}- \|\psi_n\|^4_{\ell^4(\tilde{\bTd}_n)}\Big|,$$
		  and we can conclude thanks to Lemma \ref{lpdiccont}.
                \end{proof}

                \begin{corollary}\label{immediate}
                For any
                $\delta>0$, there exist $\eta>0$ and $N_0(\delta)$,
                such that for $n>N_0(\delta)$ if $\cH_n(\psi_n) \le E_0^n(m) + \eta$
                then $\cH(\bar\psi_n) \le E_0^n(m) + \delta$.
              \end{corollary}
              \begin{proof}
                It follows immediately from Corollary \ref{energydisccont} and
                Lemma \ref{boundongrad}.
              \end{proof}

	\begin{prop} \label{convergenceofenergy}
	%Fix $m>0$, recall the definition of $E_0(m)$, $E_0^n(m)$, \eqref{Minimiztion},
	 %\eqref{discreteminimization}, then we have:
	 \begin{equation}\label{Econvcor}
		\lim_{n\to \infty} E_0^n(m) \to E_0(m).
	\end{equation} 
      \end{prop}
      
 	\begin{proof}
		Before proceeding, we emphasize the fact that all the constant 
		$c,c1,c2,c',\dots$ are independent of $n$ in this proof.\\
		Recall the definition of $Q_{m}$ as the minimizer of  \eqref{Minimiztion}. 	
		Moreover, recall the definition of the set of discrete Solitions $\cQ_m^n$, as 
		the set of mininizer of \eqref{discreteminimization}. Take $q_n \in \cQ_m^n$, 
		notice that thanks to the inequality 
		$|\psi_n(x)-\psi_n(x-1)| \geq ||\psi_n(x)|-|\psi_n(x-1)||$, we can  take 
		$q_n$ to be real-values and positive. 
		Then we have: $\cH(Q_{m})=E_0(m)$, and for all $n$, $\cH_n(q_n)=E_0^n(m)$.
                \\
		thanks to Lemma \ref{boundongrad} there exists $c>0$ uniform in $n$,
                such that $G_n(q_n) \leq c$. Therefore, we can use the 
		result of Corollary \ref{energydisccont}, and deduce that there exists $C_1$ 
		independent of $n$, such that: 
		\begin{equation} \label{convE:1}
			|\cH(\bar{q}_n)-\cH_n(q_n)| \leq \frac{C_1}{n}.
		\end{equation}		 	
			% where $\bar{q}_n$ is the linear
			% interpolation of $q_n$, see \eqref{linearinterpolation1}.\\
			For any $\psi \in H^1_{per}([0,1])$, define $\lambda_n(\psi)$ as follows: 
			\begin{equation} \label{massratio}
				\lambda_n  (\psi) = \left( \frac{m}{\cM(\psi)} \right)^{\frac12}.
			\end{equation}					 	
			% where we used the definition of $\cM$ from \eqref{conserved quantities}.
			 In particular, let $\lambda_n= \lambda(\bar{q}_n)$ and observe that 
			for $n$ sufficiently large, 
			$|\lambda_n^2-1| \leq \frac{c_0}{n}$, with $c_0$ independent of $n$,
			thanks to Lemma \ref{lpdiccont}. % (recall that we established the fact 
			% that $G_n(Q_n)$ is bounded, so we can use this Lemma)
                        More precisely, we can take
                        $c_0=\frac{2\tilde{c}}{m}$, for $n$ sufficiently large,
			where $\tilde{c}$ is given by Lemma \ref{lpdiccont}. Now, if we use 
			the definition of $\cH$, for $n$ sufficiently large we obtain:
			\begin{equation} \label{convE:2}
			\begin{split}			
			|\cH(\lambda_n\bar{q}_n)-\cH(\bar{q}_n)| &\leq |\lambda_n^2-1|\int_0^1
			\frac{|\partial_y \bar{q}_n(y)|^2}{2}dy + \frac{|\lambda_n^4-1|}{4} \int_0^1
			|\bar{q}_n(y)|^4dy \\
			  & \leq \frac{c_1}{n},
			\end{split}			
			\end{equation} 
		where $c_1$ is independent of $n$, and
		 we used the estimate $|\lambda_n^2-1| \leq \frac{c_0}{n}$; moreover, 
		 in order to treat the first
			term, we take advantage of the 
		fact that $G_n(q_n) =\int_0^1
			\frac{|\partial_y \bar{q}_n(y)|^2}{2}dy \leq c$. Lastly, the second term 
			is bounded as follows:   
                        we used the bound $\|q_n\|_{\ell^4(\bTd_n)} \leq c'$
                        (thanks to Lemma \ref{boundongrad}
		and \eqref{energylowerbound}), then we conclude by using the fact 
		$|\|q_n\|_{\ell^4(\bTd_n)}-\|\bar{q}_n\|_{L^4(\bT)}| \leq \frac{\tilde{c}'}{n}$
		which is a direct consequence of Lemma \ref{lpdiccont}.\\
		Notice that $\cM(\lambda_n \bar{q}_n)=m$; therefore, $E_0(m) \leq 
		\cH(\lambda_n \bar{q}_n)$. Combining this fact with \eqref{convE:1} and 
		\eqref{convE:2}, for $n$ large enough we have: 
		\begin{equation} \label{convE:3}
			E_0(m) \leq E_0^n(m)+\frac{c''}{n},
		\end{equation}		 	
		where $c''$ is a constant independent of $n$, and we used the fact that 
		$\cH_n(q_n)=E_0^n(m)$.\\
		
		On the other hand, recall 
		that $Q_m$ is smooth, real-valued and non-negative thanks to Theorem 
		\ref{thMinimize}.  Define 
		$Q_m^n: \bTd_n \to \bC$ as $Q_m^n(x)=Q_m(\frac{x}{n})$, 
		for $x \in \bTd_n$. Let 
		$$\tilde{\lambda}_n:=\left(\frac{m}{\cM_n(Q_m^n)}\right)^\frac12. $$ 
		Thank to the properties of $Q_m$ (in particular the fact that $Q_m$ is smooth
		with bounded $H^1$ and $L^4$ norm), for $n$ large enough we have:
		\begin{equation} \label{Econv:4}
			|\tilde{\lambda}_n^2-1| \leq \frac{c_2}{n},
		\end{equation}		  
		where one can take $c_2=\frac{4\|Q_m \|_{L^{\infty}}\|Q_m'\|_{L^{\infty}}}{m}$
		($Q'$ denotes the derivative of $Q$).			
		Moreover, since $Q_m$ is smooth, $G_n(Q_m^n)$ and $V_n(Q_m^n)$ 
		are bounded uniformly in $n$ by $\frac{\|Q'_m \|_{L^{\infty}}}{2}$, and
		$\frac{\|Q_m\|_{L^{\infty}}^4}{4}$, respectively.
		 Hence, thanks to \eqref{Econv:4} for $n$ sufficiently large we have:
		\begin{equation}\label{Econv:5}
			|\cH_n(\tilde{\lambda}_nQ_m^n)-\cH_n(Q_m^n)| \leq \frac{c_3}{n}.
		\end{equation}
		Again, since $Q_m$ is at least $C^3$, by a simple computation 
		 we get for $n$ sufficiently large:
		\begin{equation} \label{Econv:6}
			\begin{split}				
				&|\cH_n(Q_m^n)-\cH(Q_m)| \leq 
				\frac{1}{2}\sum_{x=1}^n \int_{\frac{x-1}{n}}^{\frac{x}{n}}
				\Bigg|n^2 \Big|Q_m\big(\frac{x}{n}\big) -Q_m\big(\frac{x-1}{n} \big) \Big|
				^2 -|\partial_yQ_m(y)|^2 \Bigg|dy +  \\ &\frac{1}{4}\sum_{x=1}^n 
				\int_{\frac{x-1}{n}}^{\frac{x}{n}} \Big||Q_m\big(\frac{x}{n}\big)|^4
				-|Q_m(y)|^4	\Big|dy \leq	  
				\frac{\|Q''_m\|_{L^{\infty}}\|Q'_m\|_{L^{\infty}}}{n}+ 
				\frac{\|Q'_m\|_{L^{\infty}}\|Q_m\|_{L^{\infty}}^3}{n} \\
				 & \leq \frac{c_4}{n}.
			\end{split}		
		\end{equation}			
		Therefore, combining the estimates \eqref{Econv:5} and \eqref{Econv:6}, and 
		recalling the fact that $\cH(Q_m)=E_0(m)$, we have for $n$ large enough:
		\begin{equation} \label{Econv:7}
				|\cH_n(\tilde{\lambda}Q_m^n)-\cH(Q_m)| \leq \frac{c}{n} 
				\implies E_0^n(m) \leq E_0(m) +\frac{c}{n},
		\end{equation}		 
		where we used the fact that $\cM(\tilde{\lambda}_nQ_m^n)=m$, hence 
		$E_0^n(m) \leq \cH_n(\tilde{\lambda}_nQ_m^n)$. 
		Finally, taking the limit of $n \to \infty$ in \eqref{Econv:7} and \eqref{convE:3},
		 properly ($\limsup$ and $\liminf$,
		respectively), we deduce the result
		\eqref{Econvcor}. 		
		\end{proof}
		
	We finish this section by proving the Proposition \ref{propdiscompact}:
	\begin{proof}[Proof of Proposition \ref{propdiscompact}]

          In consequence of corollary \ref{immediate} and
          Proposition \ref{Econvcor} we have that
          for any $\delta>0$, there exist $\eta>0$ and $N_0(\delta)$,
                such that for $n>N_0(\delta)$ if $\cH_n(\psi_n) \le E_0^n(m) + \eta$
                then $\cH(\bar\psi_n) \le E_0(m) + 2\delta$.
Define
$$
\lambda_{\bar{\psi_n}}=\big( \frac{m}{\cM(\bar{\psi}_n)}\big)^{\frac12} \ge 1,
$$
so that $\cM(\lambda_{\bar{\psi_n}} \bar\psi_n) = m$. Furthermore by Lemma \ref{lpdiccont}
$\lambda_{\bar{\psi_n}} \to 1$.
We also have that
\begin{equation}
  \label{eq:16}
  \begin{split}
  \cH\left( \lambda_{\bar{\psi_n}} \bar\psi_n\right) = \cH(\bar\psi_n) +
  ( \lambda_{\bar{\psi_n}}^2-1) G_n(\psi_n) -  ( \lambda_{\bar{\psi_n}}^4-1) V(\bar\psi_n)\\
  \le E_0(m) + 2\delta + ( \lambda_{\bar{\psi_n}}^2-1) C,
    \end{split}
\end{equation}
where we bounded $G_n$ thanks to Lemma \ref{boundongrad}.
By lemma \ref{lemeenergycompactness},
we have     $\|\lambda_{\bar{\psi_n}} \bar\psi_n -Q_{m}\|_{\tilde{H}^1_{per}} <\epsilon/2$,
and since
$$
\|\lambda_{\bar{\psi_n}} \bar\psi_n -\bar\psi_n\|_{\tilde{H}^1_{per}}
\le | \lambda_{\bar{\psi_n}}^2-1|^{1/2} \|\bar\psi_n\|_{\tilde{H}^1_{per}} <\epsilon/2
$$
for $n$ large enough, we conclude the proof.
          \end{proof}

	\section{Large Deviation Estimates}
        \label{sec:lde}
	% As we already discussed, we divide the proof of Theorem \ref{thlimit}, 
	% into two main steps. First, 
	In Proposition \ref{propdiscompact}, we proved that if the energy $\cH_n(\psi_n)$
        is sufficiently close to the minimal energy
	$E_n^0(m)$ for $n$ sufficiently large, then the linear interpolation
	of a configuration $\psi_n$ is close to $Q_m$ in 
	$\tilde{H}^1$-norm. In this section, we prove that
	the measure $\mu_{\beta_n,m}^n$ \eqref{gibbsmesure} 
	concentrate on configurations with minimal energy as
        $n \to \infty$, if we set $\beta_n = \vartheta(n) \beta$, where
        $\vartheta(n)$ satisfies \eqref{scaling}.
        % In this section $\beta$ and $m$ are fixed and we denote
        % $\beta_n = \vartheta(n) \beta$.
        As $m$ is fixed in this section, we will drop it from the notations.
        \\
         % Moreover, recall the definition of the sphere $S_m^n$ 
	 %        \eqref{configurationspace}, and 
	 %        the uniform probability measure on this sphere $d \mu_m^n$. 
	 %        For any measurable set 	
	 %        $A \subset S_m^n$, we denote
         %        $\mu_m^n(A)= \int_{S_m^n} \mathbbm{1}_{A}d\mu_m^n$.

	\begin{theorem} \label{gibbslimit}
		% Fix $\beta,m>0$, recall the definition of the sphere $S^m_n$
		% \eqref{configurationspace}, and the Gibbs measure $d \mu_{\beta,m}^n$ 
		% \eqref{gibbsmesure}. 
                % For any measurable set $A \subset S^m_n$,
		% denote $\mu_{\beta,m}^n(A):=\int_{S^m_n} \mathbbm{1}_{\{A\}}
          %d\mu_{\beta,m}^n$.
		% Let $\beta_n=\vartheta(n)\beta$, where $\vartheta(n)$ satisfies  	
		% \eqref{scaling}, then
          For any $\epsilon>0$, we have:
		\begin{equation} \label{gibbslimiteq}
			\lim_{n \to \infty} \mu_{\beta_n}^n( \cH_n(\psi_n)  \leq E_0^n+ \epsilon)=1,
		\end{equation}
		% where one can recall the definition of $\cH_n$, and $E_0^n(m)$ from \eqref{Ham1}, 
		% and \eqref{discreteminimization}.
	\end{theorem}

	The proof of Theorem \eqref{gibbslimit} depends on  two
        large deviation estimates for the uniform probability measure
        $d \mu_m^n$ that are proven in appendix \ref{sec:some-large-devi}:
        \begin{enumerate}
        \item 
          For any $n$
          and any $0<\mathfrak{g}$ %, and any $0<\delta<1$,
          we have: 
	\begin{equation} \label{ing:0}
	\mu_m^n(G_n(\psi_n) <\mathfrak{g}) 
	\leq \exp(-2n\ln n)\left(\frac{2\mathfrak{g}}{m}\right)^{n-1}2^n,
	\end{equation}
    %     where $\mathcal{E}_n=\frac{1}{\delta(1-\delta)^{n-1}}$, and  $\mathfrak{g}<n^{2^-}$ means 
    % there exists $\eta>0$ such that $\mathfrak{g}< n^{2-\eta}$.
    % In particular, for any
    % $c$ independent of $n$, and $n$ sufficiently large we have: 
	
    %     \begin{equation} \label{ing:1}
    %      \mu_m^n(G_n(\psi_n) <c) \leq \exp(-2n\ln n)\left(\frac{4c}{m}\right)^n
    %     \end{equation}
	This bound is proven in Lemma \ref{lemLDgradUB},
     following the same spirit as in 
	\cite{CH}, Section 10. However, because of our special scaling in $G_n$, 
	one should follow the dependence of the rate function on $n$ carefully, in 
	contrast to the estimate in \cite{CH}.
        This lemma provides the aforementioned upper bounds. 
	Combining \eqref{ing:0}  with Gagliardo-Nirenberg inequality, we can deduce 
	a suitable
	 upper bound
	for $\cH_n$. 	 
	\\
      \item
	For any $\epsilon>0$, there exists $d=d(\epsilon)$ and $N(\epsilon)$,
        such that for $n\ge N(\epsilon)$:
	\begin{equation} \label{ing:2}
	 \mu_m^n(\cH_n(\psi_n)<E_0^n +\epsilon) \geq {d}^ne^{-2n \ln 
	n}.
    \end{equation}
    This is proven in Lemma \ref{lbconditional}.

      \end{enumerate}
      
	We will proceed as follows: first, we state a proof of \eqref{gibbslimiteq}, when 
	$\beta_n= \beta n \ln n$. 
	This proof is quite simple and illustrates how does the above estimates 
	are involved. % Then we prove  \eqref{ing:1}, and \eqref{ing:2} 
	% in Lemma \ref{lemLDgrad}, and
	% \ref{lbconditional}, respectively. 
	Finally, we prove the general case $\beta_n=\vartheta(n)\beta$.

		\begin{proof}[Proof of Theorem \ref{gibbslimit} with $\beta_n=\beta n \ln n$]
			Assume $0< \epsilon<1$, % and take $\beta_n=\beta n \ln n$
		in order to prove \eqref{gibbslimiteq}, 
		it is sufficient to prove that: 
		\begin{equation}
                  \label{g0:0}
			p_n:=\frac{\int_{S^n_m} \mathbbm{1}_{\{\cH_n(\psi_n)-E^0_n \geq \epsilon \}}
			e^{-\beta_n \cH_n(\psi_n)} d \mu_m^n }
			{\int_{S^n_m} \mathbbm{1}_{\{\cH_n(\psi_n)-E^0_n < \frac{\epsilon}{2} \}}
			e^{-\beta_n \cH_n(\psi_n)} d\mu^n_m}\ \mathop{\longrightarrow}_{n\to\infty}\ 0. 
		\end{equation}
                % Take $0<\delta < \frac{\beta \epsilon}{4}$, 
                Thanks to the lower bound \eqref{ing:2}, there exists $d>0$ and $N_1$,
                such that for $n>N_1$ we have: 
		\begin{equation} \label{g0:1}
			\begin{split}			
			\bigg(
			&\int_{S^n_m} \mathbbm{1}_{\{\cH_n(\psi_n)-E^0_n < \frac{\epsilon}{2} \}}
                        e^{-\beta_n \cH_n(\psi_n)} d\mu^n_m \bigg)^{-1}\leq
                        e^{\beta_n(E_0^n+\frac{\epsilon}{2})} 
			\left[ \mu_m^n\left(\cH_n(\psi_n)-E_n^0\leq \frac{\epsilon}{2}\right) \right]^{-1} \\
                        & \leq e^{n \ln n
			\big( \beta (E_0^n+\frac{\epsilon}{2})+2\big) - n\ln d}.
			% &\leq e^{n \ln n
			% \big( \beta (E_0^n+\frac{\epsilon}{2})+2+\delta \big)}.
			\end{split}		
		\end{equation}
	Let $c> \epsilon+\frac{2}{\beta}$; recall Lemma \ref{boundongrad} and
	let $c'=C(m,c)>0$, which is given by this lemma. By using \eqref{ing:0}
	there exists $N_2$, such that for $n>N_2$: 
	\begin{equation} \label{g0:2}
		\begin{split}
				&\int_{S^n_m} \mathbbm{1}_{\{\cH_n-E^0_n(m) \geq \epsilon \}}
			e^{-\beta_n \cH_n} d \mu_m^n = 
			\\ &\int_{S^n_m} \mathbbm{1}_{\{c>\cH_n-E^0_n(m) \geq 			\epsilon \}} e^{-\beta_n \cH_n} d \mu_m^n  	
			+\int_{S^n_m} \mathbbm{1}_{\{\cH_n-E^0_n(m) \geq c \}}
			e^{-\beta_n \cH_n} d \mu_m^n 	\\
			&\leq e^{-\beta n \ln n(E_0^n(m)+\epsilon)} \mu_n^m(G_n<c')+ 
			e^{-\beta n\ln n(E^n_0(m)+c)}\\
			&\leq e^{-n \ln n\big( \beta(E^0_n(m)+ \epsilon)+2\big) +n \ln (4c'/m)}
			+e^{-\beta n \ln n (E^n_0(m)+c)} ,		
		\end{split}
	\end{equation}	 
	where in the second line we used the fact that $\{ c>\cH_n-E_0^n(m) \geq \epsilon\}
	\subset  \{G_n \leq c' \}$, thanks to the choice of $c'$, see
        Lemma \ref{boundongrad}.
      Finally, taking 
	$N>\{ N_1,N_2 \}$, and combining 
	\eqref{g0:1} and \eqref{g0:2},  gives us the following:
	\begin{equation}
		0 \leq p_n  \leq e^{-n\left(\ln n \frac{ \beta \epsilon}{2}- \ln (4c'/ d m)\right)} + 
		e^{-n \left(\ln n \left(c\beta-\frac{\beta \epsilon}{2}-2\right)- \ln d\right)}
                \ \mathop{\longrightarrow}_{n\to\infty}\ 0,
	\end{equation}	 
		thanks to the choice of $c$.		
		\end{proof}

                Now we prove Theorem \ref{gibbslimit} in the general situation with
                $\beta_n = \beta\vartheta(n)$ satisfying \eqref{scaling}% , 
	% combining  the estimates \eqref{ing:0} and 
	% \eqref{ing:2} , these estimates are proven in
	% Lemma \ref{lemLDgradUB}  and Lemma \ref{lbconditional} 
	% [\eqref{LD:ub} and \eqref{MAINlowerbound}, respectively]
        : 
	\begin{proof}[Proof of Theorem \ref{gibbslimit} with $\beta_n=\vartheta(n)\beta$]
		Fix $0< \epsilon<1$, 
		%such that $E^0_n +\frac{\epsilon}{2} \leq 0$ 
		(the 
		other cases will be straightforward).
		In order to prove \eqref{gibbslimiteq}, it is sufficient to prove \eqref{g0:0}.
                % that: 
		% \begin{equation} \label{g:0}
		% 	p_n:=\frac{\int_{S^n_m} \mathbbm{1}_{\{\cH_n-E^0_n \geq \epsilon \}}
		% 	e^{-\beta_n 
		% 	\cH_n} d \mu_m^n }
		% 	{\int_{S^n_m} \mathbbm{1}_{\{\cH_n-E^0_n < \frac{\epsilon}{2} \}}
		% 	e^{-\beta_n \cH_n} d\mu^n_m} \ \mathop{\longrightarrow}_{n\to\infty}\  0, 
		% \end{equation}
	 As for \eqref{g0:1},
		 there exist $d>0$ and $N_1$ such that for any $n>N_1$:
		\begin{equation} \label{g:1}
			\begin{split}			
			\bigg(
			\int_{S^n_m} \mathbbm{1}_{\{\cH_n-E^0_n< \frac{\epsilon}{2} \}}
			e^{-\beta_n \cH_n} d\mu^n_m \bigg)^{-1}
                        % \leq e^{\beta_n(E_0^n(m)+\frac{\epsilon}{2})} 
			% \Big( \mu_m^n(\cH_n-E_n^0(m) \leq \frac{\epsilon}{2}) \Big)^{-1} \\ 
			\leq e^{\beta
			\vartheta(n)(E_0^n +\frac{\epsilon}{2})}(d)^{-n} e^{2n \ln n}.
			\end{split}		
		\end{equation}
	Let us decompose the numerator of \eqref{g0:0} into two parts and denote them 
	by $q_n$ and $q'_n$: 
	\begin{equation} \label{g:2}
		\begin{split}					
					 \int_{S^n_m} \mathbbm{1}_{\{\cH_n-E^0_n \geq \epsilon \}}
			e^{-\beta_n \cH_n} d \mu_m^n = 
			\underbrace{\int_{S^n_m} 
			\mathbbm{1}_{\{\ln n>\cH_n-E^0_n
			\geq \epsilon \}} e^{-\beta_n \cH_n} d \mu_m^n}_{q_n}  	
			+ \\
			\underbrace{\int_{S^n_m} \mathbbm{1}_{\{\cH_n-E^0_n \geq \ln n \}}
			e^{-\beta_n \cH_n} d \mu_m^n}_{q'_n}. 	
		\end{split}	
	\end{equation}		
	We simply bound
        $q_n' \leq e^{-\beta_n(E^0_n+ \ln n)}$ and observe that:
	% thanks to \eqref{g:1}, and choice of $\epsilon$ where 
	% $E_0^n(m)+\frac{\epsilon}{2}<0$, 
        
	\begin{equation} \label{g:22}
		\bigg(\int_{S^n_m} \mathbbm{1}_{\{\cH_n-E^0_n(m) < \frac{\epsilon}{2} \}}
			e^{-\beta_n \cH_n} d\mu^n_m \bigg)^{-1}q'_n \leq 
			e^{-\ln n\left(\beta \vartheta(n) -2n\right)}
                          e^{\beta \vartheta(n) \frac{\epsilon}{2} - n \ln d} 
		   \ \mathop{\longrightarrow}_{n\to\infty}\ 0,
	\end{equation}	 
	as $n \to \infty$, where we used the fact that $d$ is a constant
	independent of $n$, as well as the condition
        $\lim_{n \to \infty}\frac{\vartheta(n)}{n} = \infty$.\\
	 Now we treat the term corresponding to $q_n$, thanks to \eqref{ing:0}. 
	  First, observe that  
	 %the following observations (they are slightly modified version of 
	 %Lemma \ref{boundongrad} and \eqref{LD:ubcorr} in Corollary \ref{cor:LDcorr}):\\
	 for any $E^0_n(m) <  a \leq n$, thanks to the inequality 
	 \eqref{GNdiscreteperiodicm}, if we have $\cH_n(\psi_n) \leq a $, we can deduce
	 $G_n(\psi_n) \leq c_1(m) +2a$, where $c_1(m)$ is a constant independent of $n$. 
	 (in fact, $c_1(m)=(\tilde{c}^2m^3+2\tilde{c}m^2)$, with $\tilde{c}=\frac{C}{4}$ 
	 and $C$ is the constant in 
	 \eqref{GNdiscreteperiodicm}); consequently, we have for any $E^0_n(m) < a \leq n$: 
	 \begin{equation} \label{g:3}
	 \mu_m^n(\cH_n \leq a) \leq \mu_m^n \big(G_n \leq 2a+c_1(m)\big).
	\end{equation}

	Recall \eqref{ing:0}:
	for any $0<\alpha<2n$ denote $\alpha_o=\frac{2\alpha}{m}$, then
	\begin{equation} \label{g:4}
		\mu_m^n \big(G_n \leq \alpha \big) \leq 2^n \alpha_o^{n-1}e^{-2 n\ln n} 
	\end{equation}		 
	holds. % , with $\cE_n = \frac{1}{\delta (1-\delta)^{n-1}}$, for any $0<\delta<1$ 
	% independent of $a$ and $n$. 
	%Notice that in order to obtain \eqref{g:4}, it is enough to argue similar to 
	%Lemma \ref{lemLDgrad}, where we replace  $\lambda=\frac{(1-\delta)n^2}{\alpha_o}$
	% in \eqref{LD:ub7}, then the rest of the argument is exactly similar to 
	 %the aforementioned Lemma). 
	 Therefore, for large $n$ thanks to \eqref{g:3}, \eqref{g:4}, we have for any
	$E^0_n(m) < a \leq n$: 
	\begin{equation} \label{g:42}
		\mu_m^n(\cH_n \leq a)  \leq  2^n e^{-2n \ln n} 
		\Big(\frac{4a+2c_1(m)}{m} \Big)^{n-1}.
	\end{equation}
%	where this bound is true for any choice of $0<\delta<1$. 	
	Take $h>0$ independent of $n$, 
	let $N= \frac{\ln n}{h}$. Then we have for $n$ sufficiently large:
	\begin{equation} \label{g:5}
	\begin{split}		
		&q_n  = \sum_{j=0}^{N-1} \int_{S^n_m} \mathbbm{1}_{
		\{E_0^n+\epsilon +jh \leq \cH_n < E_0^n+\epsilon+(j+1)h} \}
		e^{-\beta_n \cH_n} d \mu^n_m  \leq \\ &
		\sum_{j=0}^{N-1} e^{-\beta_n(E_0^n+\epsilon+jh)}
		\mu_m^n \Big( E_0^n +\epsilon+jh \leq \cH_n < E_0^n +\epsilon+(j+1)h \Big)
		\leq \\ & 2^n e^{-2n\ln n} \sum_{j=0}^{N-1} e^{-\beta_n(E_0^n+\epsilon+jh)}
		\Big(\frac{4}{m}\Big )^{n-1} 
		e^{(n-1)\ln (E_0^n+\epsilon+(j+1)h+\frac{c_1(m)}{2})}= \\
		& 2^n \Big(\frac{4}{m} \Big)^{n-1} e^{-2 n \ln n} e^{\beta_n(h+
		\frac{c_1(m)}{2})} \times \\
		&\sum_{j=1}^{N} \exp(-\beta_n(E_0^n+\epsilon+jh
		+\frac{c_1(m)}{2})+ (n-1) \ln (E_0^n+\epsilon+jh+\frac{c_1(m)}{2})),
	\end{split}		
	\end{equation}
	where we take advantage of the estimate \eqref{g:42} in the second line. 
	Notice that the term $E_0^n+
	\frac{c_1(m)}{2}>0$ thanks to the lower bound \eqref{energylowerbound}.
	Recall that $\beta_n= \vartheta(n)\beta$, with $\lim_{n \to \infty} 
	\frac{\vartheta(n)}{n} \to \infty$. Therefore, for $n$ sufficiently large 
	$-\beta_n +\frac{n-1}{x} <0$, for any $x \in [h, 2 \ln n]$. However, the later
	expression is the derivative of $-\beta_n x+ (n-1)\ln(x)$, hence, this function is 
	decreasing on the interval $[h+E_0^n+\epsilon+\frac{c_1(m)}{2},2 \ln n]$ 
	for any $n$ sufficiently large, and $-\beta_n x+ (n-1)\ln(x)$ achieves its minimum
	at $x=h+E_0^n+\epsilon+\frac{c_1(m)}{2}$ in the aforementioned interval.
	Combining this fact with \eqref{g:5} we get: 
	\begin{equation} \label{g:6}
		\begin{split}
		&q_n \leq 	2^n \Big(\frac{4}{m} \Big)^{n-1} e^{-2 n \ln n} e^{\beta_n(h+
		\frac{c_1(m)}{2})}\times \\
		 &N \exp(-\beta_n(E_0^n+\epsilon+h
		+\frac{c_1(m)}{2})+ \ln (E_0^n+\epsilon+h+\frac{c_1(m)}{2})) \\
		&= 2^n \Big(\frac{4}{m} \Big)^{n-1} e^{-2 n \ln n} 
		N\exp(-\beta_n(E_0^n+\epsilon)+ (n-1) \ln (E_0^n+\epsilon+h+\frac{c_1(m)}{2})).	
		\end{split}
	\end{equation}	 
	 Notice that $0<(E_0^n+\epsilon+h+\frac{c_1(m)}{2}) < (\epsilon+h 
	 +\frac{c_1(m)}{2}) =:c' $, and  $c'$ is a constant independent of $n$.
	 Combining the later estimate \eqref{g:6}, with \eqref{g:1} we get for $n$
	  sufficiently large: 
	 \begin{equation} \label{g:7}
	 \begin{split}
		\bigg(\int_{S^n_m} \mathbbm{1}_{\{\cH_n-E^0_n(m) < \frac{\epsilon}{2} \}}
			e^{-\beta_n \cH_n} d\mu^n_m \bigg)^{-1}q_n \leq e^{-\beta_n 
			\frac{\epsilon}{2}}\bigg( d^{-n} 2^n	(\frac{4}{m})^{n-1} (c')^{n-1}
			 \bigg) \frac{\ln n}{h} \to 0,
	 \end{split}		 
	 \end{equation}
	as $n \to \infty$. Notice that \eqref{g:7} is evident, 
	since the first term $e^{-\beta_nn }$ is super-exponentially small thanks to the 
	assumption $\beta_n = \beta \vartheta(n)$ with $\lim_{n \to \infty}
	\frac{\vartheta(n)}{n} =\infty$ and the second term is bounded by  $e^{n 
	\tilde{c}}$, where $\tilde{c}$ is a constant independent of $n$. 
	Finally, recalling the decomposition \eqref{g:2} and combining  
	\eqref{g:7} with \eqref{g:22} gives us \eqref{g0:0} and finishes the proof.  
	\end{proof}

	Finally, the proof of Theorem \ref{thlimit} is a direct consequence of Proposition
	\ref{propdiscompact}, and Theorem \ref{gibbslimit}:
	\begin{proof}[Proof of Theorem \ref{thlimit}]
		Fix $\epsilon>0$, thanks to the Proposition \eqref{thlongtimed}, in particular 
		\eqref{Longtimebehavior1d}, we have: 
		$$\lim_{n\to \infty} \lim_{t \to \infty} \mu_t^{\beta_n,n,m} 
		\Big(\|\bar{\psi}_n-Q_m\|_{\tilde{H}^1_{per}} <\epsilon \Big)=
		\lim_{n \to \infty} \mu_{\beta_n,m}^n 
		\Big(\|\bar{\psi}_n-Q_m\|_{\tilde{H}^1_{per}} <\epsilon \Big).  $$
		On the other hand let us take $\delta=\delta(\epsilon)$, which is given 
		by Proposition \ref{propdiscompact}, then for all $n>N_0(\epsilon)$ thank 
		to this proposition we have:
		\begin{equation}
			1 \geq 
			\mu_{\beta_n,m}^n 
			\Big(\|\bar{\psi}_n-Q_m\|_{\tilde{H}^1_{per}} <\epsilon  \Big)
			\geq \mu_{\beta_n,m}^n \Big(|\cH_n(\psi_n)-E_0^n(m)| 
			<\delta  \Big).			
		\end{equation}		 
		However, notice that $\lim_{n \to \infty}
		\mu_{\beta_n,m}^n \Big(|\cH_n(\psi_n)-E_0^n(m)| 
			<\delta  \Big)=1$, thanks to Theorem \ref{gibbslimit}, in particular 
			\eqref{gibbslimiteq}, and this finishes the proof
		of Theorem \ref{thlimit}, i.e.,\eqref{limitmain1}. 
	\end{proof} 	

\begin{appendices}

%	\appendix

        \section{Some large deviations for the uniform
          probability on the sphere}
\label{sec:some-large-devi}

We collect here some large deviation estimates concerning $\mu_m^n$,
the uniform probability on the complex $n$-dimensional sphere $S_m^n$,
and in particular the estimates \eqref{ing:0} and \eqref{ing:2}. 
Note that in this appendix we slightly change our notations and 
	denote the elements of $\bC^n$ by $z$ or $\underline{z}$ instead of $\psi$.
	
	\begin{lemma} \label{lemLDgradUB}
		For any $n \in \bN$, let $0<\mathfrak{g}$. For any $0<\delta<1 $ we have:
		\begin{equation} \label{LD:ub}
			\mu_m^n(G_n(\psi_n) < \mathfrak{g}) \leq 
			\frac{1}{\delta(1-\delta)^{n-1}}
                        \left(\frac{2 \mathfrak{g} }{m}\right)^{n-1} \exp(-2n\ln n).
                      \end{equation}
	\end{lemma}

	\begin{proof}
	 	
	% Let us prove the upper bound estimate \eqref{lemLDgradUB}, 
	% using the Chebyshev's inequality: 
	%	\begin{equation} 
	%		\limsup_{n \to \infty} \frac{1}{n \ln n } 
	%		\ln  \mu_m^n(G_n(\psi_n)<cn^a) \leq -(2-a).
	%	\end{equation}
	Let  $\{Z_j \} _{j=1}^{\infty}$, be a sequence of i.i.d 
	standard complex normal random variables on $(\Omega,\cF,\bP)$, 
	i.e, for any $n>0$, the probability density function of  $(Z_1,\dots, Z_n)$
	is given by:
		\begin{equation} \label{LD:ub1}
                  f(\underline{z})=\prod_{j=1}^n \frac{e^{-|z_j|^2}}{ \pi},
                  \qquad \underline{z}:=(z_1,\dots,z_n) \in \bC^n.
		\end{equation}		 
	%where $\underline{z}:=(z_1,\dots,z_n) \in \bC^n$. 
	Consequently, the random vector
	$\{\Psi_n(j)=\frac{\sqrt{m n }Z_j}{\left(\sum_\ell |Z_\ell|^2\right)^{1/2}}, \quad j=1,\dots, n\}$
        is distributed  uniformly on $S^m_n$.
	For $k \in \bTd_n$, let the random variable $\hat{Z}_k$ be defined 
	as the  Fourier transform of  $Z_1,\dots,Z_n$: 
	\begin{equation} \label{LD:ub2}
		\hat{Z}_k=\frac{1}{\sqrt{n}} \sum_{j=1}^n e^{-2\pi j \frac{k}{n}} Z_j.
	\end{equation}
	Notice that % since \eqref{LD:ub2} is unitary,
        $(\hat{Z}_1,\dots,\hat{Z}_n)$,
	has the same distribution as $(Z_1,\dots,Z_n)$. Moreover, we have the 
	following identities thanks to the properties of discrete Fourier transform: 				\begin{equation} \label{LD:ub3}
		\sum_{j=1}^n |Z_j|^2=\sum_{k=1}^n |\hat{Z}_k|^2, 
	\end{equation}
	\begin{equation} \label{LD:ub4}
		\sum_{j=1}^n |Z_j-Z_{j-1}|^2 =\sum_{k=1}^n \omega_k^2|\hat{Z}_k|^2,
	\end{equation}				
	where $\omega_k =2|\sin(\pi \frac{k}{n})|$. % First identity is true due to
	% Parseval's theorem. The second identity represents the rather well known fact
	% that discrete Fourier Transform digonlize the matrix of discrete Laplacian 
	% with periodic boundary condition.\\ 
	Denote 
	\begin{equation}
	\mathfrak{g}_o:= \frac{2 \mathfrak{g}}{n^2m},
	\end{equation}
	 and  take $0<\lambda$ such that, 
	$0<1- \mathfrak{g}_o\lambda$. By
	using Chebyshev's inequality, as well as \eqref{LD:ub3} and \eqref{LD:ub4},
	we have:   
	\begin{equation} \label{LD:ub5}
	 \begin{split}
	  &\mu^m_n(G_n(\psi_n) \leq \mathfrak{g})=% \bP(2G_n(\Psi_n)\leq 2\mathfrak{g})=
	  \bP\bigg(n^2m\frac{\sum_{j=1}^n|Z_j-Z_{j-1}|^2}{\sum_{j=1}^n|Z_j|^2} \leq 2 \mathfrak{g} 
	  \bigg)\\
	  &= \bP\Big(\sum_{k=1}^n \omega_k^2|\hat{Z}_k|^2 \leq \sum_{k=1}^n 
	  |\hat{Z}_k|^2 \mathfrak{g}_o\Big)= \bP \bigg(\exp\Big(-\sum_{k=1}^n 
	  \lambda \big(\omega_k^2-\mathfrak{g}_o \big)|\hat{Z}_k|^2 \Big) \geq 1 \bigg)
	   \\
	  &\leq \bE \bigg(\exp\Big(-\sum_{k=1}^n 
	  \lambda \big(\omega_k^2-\mathfrak{g}_o \big)|\hat{Z}_k|^2 \Big) \bigg)= 
	  \prod_{k=1}^n \bE
          \left( \exp\left(-\lambda (\omega_k^2-\mathfrak{g}_o) |\hat{Z}_1|^2\right)\right) 
		\\& = \prod_{k=1}^n \frac{1}{\lambda \big(\omega_k^2-\mathfrak{g}_o\big)+1}.
	 \end{split}
	\end{equation}		 
	Notice that in the first line, we used the fact that $\Psi_n$ is uniformly 
	distributed on $S^m_n$, and in the last line we used the fact that $\hat{Z}_k$
	are independent complex Gaussian variable with the same distribution as $Z_i$, 
	as well as the choice of $\lambda$, which permits us to compute the last  
	expectation. We emphasize the fact that the last bound holds for any 
	$0< \lambda<\mathfrak{g}_o^{-1} =\frac{n^{2}m}{2 \mathfrak{g}}$, 
	which can depend on $n$. In fact, our choice 
	of $\lambda$ depends on $n$. \\
	Before proceeding, let us recall the following trigonometric identity:
%	(\textcolor{red}{Reference?, proof in appendix?}):
		\begin{equation} \label{LD:ub6}
			\prod_{k=1}^{n-1} \sin(\frac{\pi k}{n})= \frac{n}{2^{n-1}}, \implies
			\frac{1}{\prod_{k=1}^{n-1} \omega_k^2} = \frac{1}{n^2}.
		\end{equation}
                For any  $0<\delta<1$, let us take
                $\lambda=\frac{(1-\delta)}{\mathfrak{g}_o}$.
                Notice that we have $1-\lambda \mathfrak{g}_o =\delta$.
                Thanks to the choice of $\lambda$, by using 
	\eqref{LD:ub5} and \eqref{LD:ub6}, we obtain 
	\begin{equation} \label{LD:ub7}
		\begin{split}		
		\mu^m_n(G_n(\psi_n) < \mathfrak{g}) 
		\leq \prod_{k=1}^n \frac{1}{\lambda \big(\omega_k^2-\mathfrak{g}_o\big)+1}
                \leq \frac{1}{\delta }  \prod_{k=1}^{n-1}
                \frac{1}{\lambda \omega_k^2}
		\leq \frac{1}{\delta} \frac{1}{(1-\delta)^{n-1}}
		\mathfrak{g}_o^{n-1}  \frac{1}{n^2} \\
		=\frac{1}{\delta(1-\delta)^{n-1}} \exp(-2n \ln n) \left(\frac{2 
		\mathfrak{g}}{m} \right)^{n-1}.
		\end{split}	
	\end{equation}	  
%	This finishes the proof of \eqref{LD:ub} and Lemma \ref{lemLDgradUB}.
      \end{proof}

      Notice that the bound \eqref{ing:0} corresponds to the choice $\delta = 1/2$.
	
	We obtain now the lower bound \eqref{ing:2}, 
	indicating that 
	set of configurations with close to minimal energy is "large enough".\\
		
	\begin{lemma} \label{lbconditional}
	%As usual recall the definition of $S^n_m$, and $d\mu^n_m$ as
	%the uniform probability measure on $S^n_m$. 
		  For any $\epsilon>0$, there exist $N(\epsilon)$, and 
		  a constant $\mathrm{c}=\mathrm{c}(\epsilon)$,  
		  independent of $n$, such that for $n>N(\epsilon)$ we have: 
		  \begin{equation} \label{MAINlowerbound}
		  \mu^n_m(\cH_n < E_0^n + \epsilon) \geq \mathrm{c}^n e^{-2 n \ln n}.
		\end{equation}		   
	\end{lemma} 
  \begin{proof}

    Denote by $Q$ a discrete Soliton, we have that
    $E_0^n = \cH_n(Q) = G_n(Q) - V_n(Q)$. we know from the results of section \ref{Dissol}
    that $Q$ is uniformly bounded in $n$ as well as $G_n(Q)$ and  $V_n(Q)$.
    Observe that
    \begin{equation}
      \label{eq:2}
      \begin{split}
        &\left\{ \psi \in S^n_m : \cH_n(\psi) < E_0^n + \epsilon\right\}\ \supset \\
        &\quad \left\{\psi \in S^n_m: |G_n(\psi) - G_n(Q)| \le \epsilon/2
          ,  |V_n(\psi) - V_n(Q)| \le \epsilon/2 \right\}.
      \end{split}
    \end{equation}
   Consequently, we need to construct a neighborhood
    $\tilde{\mathcal A} \subset S^n_m $ of $Q$
    that is contained in the set on the RHS of \eqref{eq:2}, and such that
    $\mu^n_m(\tilde{\mathcal A}) \ge \mathrm{c}^n e^{-2 n \ln n}$
    for some constant depending on $\varepsilon$.

    % Since $Q$ is uniformly bounded, it is easy to see that if
    % \begin{equation}
    %   \label{eq:3V}
    %   \sup_j |\psi_j - Q_j|\le \frac{\epsilon}n \qquad \Longrightarrow
    %   |V_n(\psi) - V_n(Q)| \le \epsilon.
    % \end{equation}
    % and
    % \begin{equation}
    %  \label{eq:3G}
    %   \sup_j |\psi_j - Q_j|\le \frac{\epsilon}n \qquad \Longrightarrow
    %   |G_n(\psi) - G_n(Q)| \le \epsilon^2.
    % \end{equation}

    % So it is enough to prove that
    % \begin{equation}
    %   \label{eq:4}
    %   \mu^n_m \left(\sup_j |\psi_j - Q_j|\le \frac{\epsilon}n\right) \ge  \mathrm{c}^n e^{-2 n \ln n}.
    % \end{equation}
    % Since the volume of this set does not depend on $Q$ and $ \mu^n_m $ is uniform,
    % the LHS is equal to $\mu^n_m \left(\sup_j |\psi_j - \sqrt{m/2} (1+i)|\le \frac{\epsilon}n\right)$
    % and the lower bound follow by the following argument.
    
    % The set $\{\sup_j |\psi_j -  Q_j|\le \frac{\epsilon}n\} \subset S^n_m$ contains the
    % hyperspherical cap
    % $\{\sum_j |\psi_j -  Q_j|^2 \le \frac{\epsilon^2}{n^2}\} \subset S^n_m$. The surface of
    % this  hyperspherical cap is larger that the surface of the real sphere  $S^{2n-1}_{\epsilon/2n}$.
    % The ratio of the surface of $S^{2n-1}_{\epsilon/2n}$ with $S^{2n}_m$
    % gives a lower bound of order $c^n e^{-3n\ln n}$ that is too small.
    % So we need to construct a larger neighborhood of $Q$.

    Let us identify $\mathbb C^n \sim \mathbb R^{2n}$, and
    denote the corresponding real components of $Q$ by $(q_1, \dots,q_{2n})$, and the
    components of a generic $\psi \in S^n_m \sim \mathbb S^{2n}_{\sqrt{nm}}$
    as $(x_1, \dots, x_{2n})$. We can choose the discrete Soliton $Q$, such that
    $q_{2n} \ge q_j \ge 0$.

For any small $\delta>0$, define the set $\tilde{A}_{\delta} \subset \bR^{2n-1}$ as follows:
	\begin{equation} \label{LD:cond4s}
		\tilde{A}_{\delta} = \bigg \{ \underline{\xi} \in \Big[-\frac{\delta}
		{2n}, \frac{\delta}{2n} \Big]^{2n-1} \bigg| \quad  
		\Big|\sum_{j=1}^{2n-1} \xi_j q_j \Big| \leq
		\frac{q_{2n} \delta}{2\sqrt{n}} \bigg \}.
	\end{equation}	    	   
The volume of this set can be easily estimated by
\begin{equation}
  \label{eq:3}
  \text{vol}(\tilde{A}_{\delta}) \ge \frac 13 \left(\frac{\delta}{n}\right)^{2n-1}.
\end{equation}
We postpone the proof of \eqref{eq:3} later.

 We now define our neighborhood of $Q$ as
\begin{equation}
  \label{eq:5}
  \tilde{\mathcal A}_\delta =
  \left\{  \underline{x} \in \mathbb S^{2n}_{\sqrt{nm}}: x_j = q_j + \xi_j, j=1\dots,2n-1;\ 
     \underline{\xi} \in \tilde{A}_{\delta}\right\}.
 \end{equation}
 Notice that if $\underline{x} \in \tilde{\mathcal A}_\delta$, we have automatically that
 $x_{2n} = (nm- \sum_{j=1}^{2n-1}(q_j + \xi_j)^2)^{\frac12}$. Furthermore,
 we have that
 \begin{equation}
   \label{eq:7}
   |x_{2n} - q_{2n}| \le \frac {2\delta}{\sqrt n}.
 \end{equation}

 It is easy to check that if $\underline{x} \in \tilde{\mathcal A}_\delta$ then
 $ |V_n(\underline{x}) - V_n(Q)| \le 2c\delta/n$, where $c$ is a constant independent of $n$.
 About the gradients term, denoting $\xi_{2n} = x_{2n} - q_{2n}$, we have
 \begin{equation}
   \label{eq:15}
   \begin{split}
     G_n(\underline{\xi}) &= \frac n2 \sum_{j=1}^{n-1} (\xi_{i+1}- \xi_i)^2 + \frac n2 (\xi_{n}- \xi_1)^2\\
     &+  \frac n2 \sum_{j=1}^{n-2} (\xi_{n+i+1}- \xi_{n+i})^2 + \frac n2 (\xi_{2n}- \xi_{2n-1})^2
       + \frac n2 (\xi_{2n}- \xi_{n+1})^2
       \le 6 \delta^2,
     \end{split}
     \end{equation}
     and
 \begin{equation}
   \label{eq:6}
   \begin{split}
     \left| G_n(\underline{x} ) - G_n(Q)\right| =
     \left| n \sum_j (q_j - q_{j-1})(\xi_j - \xi_{j-1}) + G_n(\underline{\xi})\right|\\
     \le \left(2G_n(Q)\right)^{1/2} \left(2G_n(\underline{\xi})\right)^{1/2} + G_n(\underline{\xi})\\
     \le C \left(2G_n(\underline{\xi})\right)^{1/2} + G_n(\underline{\xi}) \le C' \delta.
   \end{split}
 \end{equation}
 It follows that, choosing $\delta< \epsilon/2C'$, the set $\tilde{\mathcal A}_\delta$
 is contained in the set defined in \eqref{eq:2}.

In order to compute $\mu_m^n( \tilde{\mathcal A}_\delta)$ we use the following
change of variable formula  for any measurable $f:\bS^{2n}_r \to \bR$: (cf.
 	Appendix  A of \cite{integral})
 	 \begin{equation} \label{LD:cond7s}
           \int_{\bS_r^{2n}} f(\underline{x}) d\sigma_r^{2n}(\underline{x})
           = \frac{r^2}{2r^{2n}nV(\bB^{2n}_1)} 
           \int_{\bB^{2n-1}_r} \frac{f(\underline{y},\sqrt{r^2-\|\underline{y}^2\|})
             +f(\underline{y},-\sqrt{r^2-\|\underline{y}^2\|})}
 	 	{\sqrt{r^2-\|\underline{y}\|^2}}dy_1 \dots dy_{2n-1},
 	 \end{equation}
         where $\|.\|$ denotes the Euclidean norm in $\bR^{2n-1}$, and
         $V(\bB^{2n}_1) = \frac{\pi^n}{n!}$ denotes
 	the volume of the unit ball.
        Applying the above formula and noticing that $nm \ge \sqrt{nm - \|\underline{y}\|^2}$,
        we have
        \begin{equation}
          \label{eq:8}
          \begin{split}
            \mu_m^n( \tilde{\mathcal A}_\delta) = \frac{m n!}{2(nm)^{n} \pi^n}
            \int_{\bB^{2n-1}_{\sqrt{nm}}}
            \frac{ \mathbbm{1}_{\tilde{\mathcal A}_\delta}(\underline{y}, \sqrt{nm-\|\underline{y}\|^2})
              + \mathbbm{1}_{\tilde{\mathcal A}_\delta}(\underline{y}, -\sqrt{nm-\|\underline{y}\|^2})}
            {\sqrt{nm - \|\underline{y}\|^2}} dy_1 \dots dy_{2n-1}\\
            \ge  \frac{ n!}{2n(nm)^{n} \pi^n}
             \int_{\bB^{2n-1}_{\sqrt{nm}}}
     \left[\mathbbm{1}_{\tilde{\mathcal A}_\delta}(\underline{y}, \sqrt{nm-\|\underline{y}\|^2})
              + \mathbbm{1}_{\tilde{\mathcal A}_\delta}(\underline{y}, -\sqrt{nm-\|\underline{y}\|^2})\right]
            dy_1 \dots dy_{2n-1}\\
            = \frac{ n!}{n(nm)^{n} \pi^n} \int_{-\frac{\delta}{n}}^{\frac{\delta}{n}}\dots
            \int_{-\frac{\delta}{n}}^{\frac{\epsilon}{n}}
            \mathbbm{1}_{\tilde{A}_{\delta}}(\underline{\xi}) d\xi_1 \dots d\xi_{2n-1}
            \ge  \frac{ n!}{n(nm)^{n} \pi^n} \frac 13 \left(\frac{\delta}{n}\right)^{2n-1}
          \end{split}
        \end{equation}
and by Stirling approximation we have the desired lower bound.

\end{proof}

\begin{proof}[Proof of \eqref{eq:3}]

  Let 
	  $\{\xi_j \}_{j=1}^{\infty}$ be a sequence of i.i.d random variables
          uniformly distributed on $[-\frac{\delta}{2n}, \frac{\delta}{2n}]$.
          Thanks to Chebyshev's inequality we get: 
	  \begin{equation} \label{LD:cond3}
	  	\begin{split}
	  	 & \bP \bigg(\Big|\sum_{j=1}^{2n-1} \xi_j q_j \Big| \leq
		  \frac{q_{2n}\delta}{2\sqrt{n}} \bigg) =
		1-\bP \bigg(\Big|\sum_{j=1}^{2n-1} \xi_j q_j \Big|^2 >
		  \Big(\frac{q_{2n}\delta}{2\sqrt{n}}\Big)^2 \bigg) \geq \\ 
		&1-\frac{4 n}{q_{2n}^2 \delta^2}
		\bE \bigg(\Big|\sum_{j=1}^{2n-1} \xi_j q_j \Big|^2\bigg)\geq
                1- \frac{4n}{\delta^2}\bE(\xi_1^2)
                \sum_{j=1}^{2n-1} \left(\frac{q_j}{q_{2n}}\right)^2
                = 1 - \frac{1}{3n}\sum_{j=1}^{2n-1} \left(\frac{q_j}{q_{2n}}\right)^2
                \geq	1-\frac{2}{3},
		\end{split}	  
              \end{equation}
              where we used our choice of the discrete Soliton $0\le q_j\le q_{2n}$.
\end{proof}

We conclude this section mentioning some more precise
limits on the large deviations for the
uniform measure on the sphere,
with a matching lower bound for Large deviation estimate \eqref{LD:ub}.
\textit{These results are not used for proving theorem \ref{gibbslimit}
  and theorem \ref{thlimit}, so their proof would be published in
  a future work  \cite{amir-so2}.}
For $0\le a <2$ we have:
	\begin{equation} \label{LD:ubcorr}
		\limsup_{n \to \infty}\frac{1}{n} \ln(e^{(2-a)n \ln n} 
		\mu_m^n(G_n < cn^a)) \leq \ln(\frac{2c}{m}). 
	\end{equation}	  
	% Moreover, the expressions \eqref{LD:lb17}, \eqref{LD:lb3} and \eqref{LD:lb13}
	%   give us the same correction for the lower bound \eqref{LD:lb}: 
	\begin{equation} \label{LD:lbcorr}
		\liminf_{n \to \infty} \frac{1}{n} \ln (e^{(2-a)n \ln n} 
		\mu_m^n(G_n\leq cn^a)) \geq \ln(\frac{2c}{m}).
	\end{equation}			

	\section{Hypoellipticity}\label{app:hor}
        
	In this section, we prove that the generator  \eqref{generatord}, is hypoelliptic, 
	and therefore the invariant measure has a smooth density. Notice that we add the 
	subscript $n$, to emphasize the dependence on $n$.
	\begin{lemma} \label{hypolem}
		 Recall the operator $L_n= A_n+S_n$, 
		where  $A_n=\cA$ and $S_n=\cS$ are given by 
		\eqref{generatorhamd}, \eqref{generatorrandd}, respectively. Then $L_n$ is
		hypoelliptic. Consequently, the invariant measure has smooth density with respect
		to $d\mu ^m_n$.	
	\end{lemma} 
	\begin{proof}
	Let us the fix the parameters $h=s=\gamma=1$, the proof for other cases is similar.   		
	In order to prove this lemma, it is sufficient to show that $L_n$ satisfies 
	the so-called H\"{o}rmander's condition. Then the hypoellipticity, and smoothness
	of the invariant measure follow by the H\"{o}rmander's Theorem (hypoellipticity 
	follows from Thorem 22.2.1 of \cite{hormander}, for a general review
	one can also see  \cite{bellet}, and \cite{hairer}).
	We prove this condition in the case $d=1$ in details, the generalization to  higher
	dimensions  is a matter of messier algebra (We comment on this at the end of 
	the proof).
	Let us denote $Y_0=A_n$ 
	and $Y_x=\partial_{\theta(x)}$ for $x \in \bTd_n$. $L_n$ satisfies the 
	H\"{o}rmander's condition if the Lie algebra generated by the family
	$$\{Y_x \}_{x=1}^n, \quad \{[Y_x,Y_y]\}_{x,y=0}^n, \quad 
	\{[[Y_x,Y_y],Y_z]\}_{x,y,z=0}^n,\dots, $$
	has full rank (here $2n-1$) at every point $\psi \in S^m_n$. \\
	Let us define the following notation: for $x,y \in \bTd_n$ and symbols $i,r$,
	we define $\cR_{x^i,y^r}$, $\cR_{x^i,y^i}$, $\cR_{x^r,y^r}$, and $\cR_{x^r,y^i}$ as
	 the following rotations: 
	\begin{equation} \label{rotation}
	\begin{split}
		&\cR_{x^r,y^r} = \psi_r(x)\partial_{\psi_r(y)} -\psi_r(y) 
		\partial_{\psi_r(x)},\quad 
		\cR_{x^r,y^i} = \psi_r(x)\partial_{\psi_i(y)} -\psi_i(y)
		\partial_{\psi_r(x)},\\	
		&\cR_{x^i,y^r} = \psi_i(x)\partial_{\psi_r(y)} -\psi_r(y) \partial_{\psi_i(x)},
		\quad 
		\cR_{x^i,y^i} = \psi_i(x)\partial_{\psi_i(y)} -\psi_i(y) \partial_{\psi_i(x)}.
	\end{split}
	\end{equation}
	We can rewrite $\partial_{\theta(x)}$, and the Hamiltonian operator $A_n$ 
	in terms of these
	rotations: 
	\begin{equation}  \label{com0}
		\begin{split}	
			&\partial_{\theta(x)}=\cR_{x^r,x^i},\\
			&A_n= \sum_{x \in \bTd_n} 
			\cR_{(x+1)^r,x^i} +\cR_{x^r,(x+1)^i} -2 \cR_{x^r,x^i} +
			\kappa|\psi(x)|^{p-1}\cR_{x^r,x^i}.
		\end{split}
	\end{equation}
	Observe that for any $\alpha_1,\alpha_2,\alpha_3,\alpha_4 \in 
	\{x^{\mu} | x \in \bTd_n, \mu \in 
	\{ {r,i} \}  \}$, 
	(these indices are of the form $x^i$,$x^r$), we have (recall $[a,b]=ab-ba$): 
	\begin{equation} \label{com2}
		\begin{split}		
		[\cR_{\alpha_1,\alpha_2},\cR_{\alpha_3,\alpha_4}]= 
		&\delta_{\alpha_1,\alpha_4} \cR_{\alpha_2,\alpha_3}+
		\delta_{\alpha_2,\alpha_3} \cR_{\alpha_1,\alpha_4}- 
		\delta_{\alpha_1,\alpha_3} \cR_{\alpha_2,\alpha_4}
		-\delta_{\alpha_2,\alpha_4} \cR_{\alpha_1,\alpha_3}=\\
		&\sum_{i,j=1}^4 \delta_{\{i+j-5\}} \cR_{\alpha_i \alpha_j}\delta_{\alpha_k,
		\alpha_l}(-1)^{i+j+1},
		\end{split}	
	\end{equation}
	where $\{k,l\} := \{1,2,3,4 \}  \setminus \{i,j \}$.\\
	We rewrite the following commutators in terms of these rotations for every 
	$x \in \bTd_n $: 
	\begin{equation} \label{com3}
			\begin{split}		
			&\partial_{\theta(x)}=\cR_{x^r,x^i}, \\
		&\cA_x := [A_n,\partial_{\theta(x)}]=\cR_{x^r,(x+1)^r}+\cR_{x^i,(x+1)^i}-
		\cR_{(x-1)^r,x^r}
		-\cR_{(x-1)^i,x^i}, \\
		&\cA _{x,x+1}:=
		[[A_n,\partial_{\theta(x)}],\partial_{\theta(x+1)}]= \cR_{x^r,(x+1)^i}-
		\cR_{x^i,(x+1)^r},\\
		&\cA_{x,x+1,x}:=
		[[[A_n,\partial_{\theta(x)}],\partial_{\theta(x+1)}],\partial_{\theta(x)}]=
		\cR_{x^r,(x+1)^r}+\cR_{x^i,(x+1)^i}.
		\end{split}	
	\end{equation}
	We can compute the following commutators: 
	$\cA^{(2)}_{x,x+1}:=[\cA_{x,x+1},\cA_{x+1,x+2,x+1}]$ and \\
	$[\cA^{(2)}_{x,x+1},\partial_{\theta(x+2)}]$ thanks to \eqref{com2}, and observe that
	$\cR_{x^r,(x+2)^i}-\cR_{x^i,(x+2)^r}$ and $\cR_{x^r,(x+2)^r}+\cR_{x^i,(x+2)^i}$ 
	belong to our Lie algebra. Repeating this process, following an induction, we observe 
	that for $x,y \in \bTd_n$ the following terms are in the Lie algebra generated by 
	$\{Y_x\}_{x=1}^n$, $\{[Y_x,Y_y]\}_{x,y=0}^n$, 
	$\{[[Y_x,Y_y],Y_z]\}_{x,y,z=0}^n$,$\dots$:
	\begin{equation} \label{com4}
		\cG^o_n:=	\{ \cR_{x^r,x^i},\cR_{x^r,y^i}-\cR_{x^i,y^r} ,
		\cR_{x^r,y^r} + \cR_{x^i,y^i} 
		\big| x,y \in \bTd_n\}.
	\end{equation}	 		 
	Notice that in the linear case (absence of non-linearity i.e., $p=2$), the terms 
	appeared in \eqref{com4} represent a basis for the Lie algebra. (All the 
	elements are linear combination of these terms). \\
	In the following, we observe that $\cG^o_n$ \eqref{com4}, has rank $2n-1$  for any
	$\psi \in S^n= \{ \psi \in \bC^n|  \sum_{x=1}^n |\psi(x)|^2=1  \}$, notice
	that we consider $S^n$ as a $2n-1$ real sphere $\bS^{2n-1}$ (the case where we 
	replace $S^n$ by $S^n_m$ can be treated similarly). Let us  
	proof by an induction.  The case $n=1$ is trivial, since 
	$\partial_{\theta_1}=\psi_r(1)\partial_{\psi_i(1)}-\psi_i(1)\partial_{\psi_r(1)}$
	has rank one for any $\psi(1) \in \bS^1$ ($|\psi_r(1)|^2+|\psi_i(1)|^2=1$).\\ 
	Assume $\cG_n^o$ has rank $2n-1$ at every point of $\bS^{2n-1}$, we prove that
	$\cG_{n+1}^o$ has rank $2n+1$ at every point of $\bS^{2n+1}$. We split the proof
	into two cases:\\
	\textit{Case 1.}\\
	 Take $\psi \in S^{n+1}$, 
	 and assume that there exists at least one point $x$, such that $|\psi(x)|=0$, we can 
	 take $x=n+1$, since we are in the periodic setup. We have 
	 $\psi_r(n+1)=\psi_i(n+1)=0$; therefore, $\hat{\psi}=(\psi_1,\dots,\psi_n) \in S^n$, 
	 and $\cG_n^o$ has rank $2n-1$ by induction hypothesis. On the other hand, since 
	 $\hat{\psi} \in S^n$,  there exists $y \in \bTd_n$, such that $|\psi(y)|\neq 0$. First,
	 observe that 
	 \begin{equation}
	 	\begin{split}	
 		B:= &\{ \psi_r(y)\partial_{\psi_i(n+1)}-\psi_i(y)\partial_{\psi_r(n+1)}, 
	  \psi_r(y)\partial_{\psi_r(n+1)}+\psi_i(y)\partial_{\psi_i(n+1)}\} =
	  \\ 
	  &\{\cR_{y^r,(n+1)^i}-\cR_{y^i,(n+1)^r},\cR_{y^r,(n+1)^r}-\cR_{y^i,(n+1)^i}\}
	  \subset \cG_{n+1}^o  ,
	  \end{split}
	  \end{equation}
 	 has rank two (this is straightforward, since $(\psi_r(y),\psi_i(y))\neq 0$, and one
	 can see a linear combination of elements of $B$ is zero iff  $|\psi(y)|=0$). Then 
	 the result follows from the induction hypothesis, as well as the fact that $B$ is 
	 orthogonal to $\cG_n^o$.\\
	\textit{Case2.}\\
	Take $\psi \in S^{n+1}$ and assume $|\psi(x)| \neq 0$ for all $x \in \bTd_{n+1}$.
	In this case,  we claim the set 
	\begin{equation}
		\cG^1_{n+1}:=\{\cR_{(n+1)^r,(n+1)^i},\cR_{(n+1)^r,x^i}-\cR_{(n+1)^i,x^r},
		\cR_{(n+1)^r,x^r}+\cR_{(n+1)^i,x^i} \big| x \in \bTd_n \} \subset \cG_n^o,  
	\end{equation}	    
	has rank $2n+1$. In fact, this set has $2n+1$ elements, where we observe that 
	they are linearly independent. Take real \footnote{Notice that we are considering
	the Field $\bR$ here, by decomposing $\psi$ into real and imaginary parts.}
	 coefficients $\{a_x,b_x,c\}_{x=1}^n$ 
	such that 
	$$c \cR_{(n+1)^r,(n+1)^i} + \sum_{x=1}^n a_x\Big(\cR_{(n+1)^r,x^i}-\cR_{(n+1)^i,x^r}
	\Big) + b_x \Big(\cR_{(n+1)^r,x^r}+\cR_{(n+1)^i,x^i} \Big) =0. $$
	Computing the coefficients of $\partial_{\psi_r(x)}$ and $\partial_{\psi_i(x)}$, 
	for any $x \in \bTd_n$ we get: 
	\begin{equation} \label{com6}
		\begin{split}
			&\Big(  b_x \psi_r(n+1) -a_x \psi_i(n+1) \Big) \partial_{\psi_r(x)}=0,\\
			&\Big(a_x \psi_r(n+1) + b_x  \psi_i(n+1) \Big) \partial_{\psi_i(x)}=0.		
		\end{split}	
	\end{equation}
	Notice that if $(a_x,b_x) \neq 0$, then   
	$\det \Big(\begin{matrix}
	b_x & -a_x \\  a_x & b_x
	\end{matrix} \Big) > 0$. 
	However, in order to \eqref{com6} holds, the later cannot happen, 
	since $(\psi_r(n+1),\psi_i(n+1)) \neq 0$; therefore 
	$a_x=b_x=0$ for all $x \in \bTd_n$, and we can deduce $c=0$, which yields the result 
	in the case $d=1$. \\
	In order to prove the result for $d>1$, for any $x,y \in \bTd_n^d$, and
	any $\mu, \nu \in \{r,i \}$,  we  define $\cR_{x^{\mu},y^{\nu}}$, similar to 
	\eqref{rotation}. Recall $\{e_j \}_{j=1}^d$ as the canonical basis of $\bR^d$,
	then \eqref{com0} will be modified as:
	\begin{equation}  \label{com0d}
		\begin{split}	
			&\partial_{\theta(x)}=\cR_{x^r,x^i},\\
			&A_n= \sum_{x \in \bTd_n} \sum_{j=1}^d 
			\left(\cR_{(x+e_j)^r,x^i} +\cR_{x^r,(x+e_j)^i} -2 \cR_{x^r,x^i} \right) +
			 \sum_{x \in \bTd_n} \kappa|\psi(x)|^{p-1}\cR_{x^r,x^i}.
		\end{split}
	\end{equation}
	 The identity \eqref{com2} remains true by taking 
	 $\alpha_1,\alpha_2,\alpha_3,\alpha_4 \in 
	\{x^{\mu} | x \in \bTd_n^d, \mu \in 
	\{ {r,i} \}  \}$. This leads to
	the  following modification of \eqref{com3}, for any $x \in \bTd_n^d$, and $1\leq k 
	 \leq d$:
	\begin{equation} \label{com3d}
			\begin{split}		
			&\cA_x := [A_n,\partial_{\theta(x)}]= \sum_{j=1}^d
			\cR_{x^r,(x+e_j)^r}+\cR_{x^i,(x+e_j)^i}-
		\cR_{(x-e_j)^r,x^r}
		-\cR_{(x-e_j)^i,x^i}, \\
		&\cA _{x,x+e_k}:=
		[[A_n,\partial_{\theta(x)}],\partial_{\theta(x+e_k)}]= \cR_{x^r,(x+e_k)^i}-
		\cR_{x^i,(x+e_k)^r},\\
		&\cA_{x,x+e_k,x}:=
		[[[A_n,\partial_{\theta(x)}],\partial_{\theta(x+e_k)}],\partial_{\theta(x)}]=
		\cR_{x^r,(x+e_k)^r}+\cR_{x^i,(x+e_k)^i}.
		\end{split}	
	\end{equation}
	Following the exact same strategy as in the previous case, by an induction 
	we observe that all the terms of the form 
	$\cR_{x^r,(x+l_ke_k)^i}-\cR_{x^i,(x+l_ke_k)^r}$ and $\cR_{x^r,(x+l_ke_k)^r}+
	\cR_{x^i,(x+l_ke_k)^i}$, for any $x \in \bTd_n^d$, any $1 \leq k \leq d$, and
	any $l_k \in \bTd_n$, belong to our Lie algebra. 
	Notice that thanks to \eqref{com2}, we have: 
	 	\begin{equation}
	 		\begin{split}
	 		&[\cR_{x^r,(x+l_ke_k)^i}-\cR_{x^i,(x+l_ke_k)^r},
	 		\cR_{(x+l_ke_k)^r,(x+l_ke_k+l_{k'}e_{k'})^i}-\cR_{(x+l_ke_k)^i,
	 		(x+l_ke_k+l_{k'}e_{k'})}]= \\
	 		&-\left(\cR_{x^r,(x+l_ke_k+l_{k'}e_{k'})^r}
	 		+\cR_{x^i,(x+l_ke_k+l_{k'}e_{k'})^i}\right)\\
	 		&[\cR_{x^r,(x+l_ke_k+l_{k'}e_{k'})^r}
	 		+\cR_{x^i,(x+l_ke_k+l_{k'}e_{k'})^i}, \cR_{x^r,x^i}]=
	 		\cR_{x^i,(x+l_ke_k+l_{k'}e_{k'})^r}
	 		-\cR_{x^i,(x+l_ke_k+l_{k'}e_{k'})^r}.
			\end{split}	 	
	 	\end{equation}
			Repeating the above procedure for $d-1$ times, we can deduce the following set
			is included in our Lie algebra:
			\begin{equation} \label{com4d}
		\cG^{o,d}_n:=	\{ \cR_{x^r,x^i},\cR_{x^r,y^i}-\cR_{x^i,y^r} ,
		\cR_{x^r,y^r} + \cR_{x^i,y^i} 
		| x,y \in \bTd_n^d\}.
	\end{equation}
		Recall that we observed that the rank of $\cG^o_n$ is $2n-1$. However, 
		due to symmetry one can observe that $\cG_{n^d}^o$ and $\cG^{o,d}_n$ has the same 
		rank and this finishes the proof. 	
	\end{proof}
		
	\begin{rem} \label{hypormk}
		Notice that the proof of Lemma \ref{hypolem} can be adapted to any other 
		non-linearity of the form $f(|\psi(x)|)$, where $f$ is smooth.
	\end{rem}

\section{Discrete Gagliardo-Nirenberg  Inequality} \label{secGN}
 	We present different versions of the Gagliardo-Nirenberg inequality. 
 	This inequality is crucial in the study of the sub-critical nonlinear focusing
 	Schr\"{o}dinger  equation, for proving the well-posedness and characterization 
 	of the Solitons (cf. \cite{cazenave03},\cite{raphael},\cite{tao06},\cite{GLT}). In 
 	particular, this inequality 
 	has been used in the
 	the proof of Theorem \ref{thMinimize} in \cite{GLT}. We take advantage  of 
 	the discrete version of this inequality, so we can establish properties of
 	configurations with minimal or close to minimal energy. 
\\
	Gagliardo-Nirenberg inequality states that for every $u \in H^1(\mathbb{R}^d)$,  
	and $1<p<1+\frac4d$, there exists a constant $C(p,d)$, such that 
	(cf. \cite{cazenave03},\cite{raphael},\cite{tao06}):
		\begin{equation}\label{GNcontinous}
			\|u\|_{L^{p+1}}^{p+1}= \int |u|^{p+1} \leq C(d,p) \Big(\int |\nabla u|
			^2\Big)^{\frac{d(p-1)}{4}} \Big(\int |u|^2\Big)^{\frac{p+1}{2}-\frac{d(p-1)}{4}}.
		\end{equation}
	While we are focusing on the case where $d=1$, and $p=3<1+\frac4d=5$ and 
	the domain is periodic, we state the following version from [\cite{GLT} Section3.2, 
	\cite{LRS88}
	Lemma 4.1].
	For all $u \in H^1_{per}= H^1(\bT)$, there exists a constant $C>0$:
		\begin{equation} \label{GNperiodic}
			\begin{split}	
	\|u\|_{L^4}^4 &=\int_0^1 |u|^4 \leq C(\|\partial_xu\|_{L^2}\|u\|_{L^2}^3+
	\|u\|_{L^2}^4)  \\
	&=C\left(\Big(\int_0^1|u|^2\Big)^{\frac32}\Big(\int_0^1|\partial_xu|^2\Big)^{\frac12}+
	\Big(\int_0^1|u|^2\Big)^2 \right).
			\end{split}
		\end{equation}

	We need counterparts of these inequalities in the discrete setting, in order to 
	obtain these inequalities, we generalize results from [\cite{CH} section 17]. 
	First, we define: Fix $n>0$, and consider a function 
	$f:\bTd_n \to \mathbb{C}$, define the discrete $\ell^p(\tilde{\bTd_n})$ norm  
	of $f$, for $p \geq 1$, as:

		\begin{equation} \label{discretelpnorm}
			\|f\|_{\ell^p(\bTd_n)}=
			\Big(\frac{1}{n}\sum_{x \in \bTd_n} |f(x)|^{p}
			\Big)^{\frac{1}{p}}.
		\end{equation}  
	Notice that our definition differs from the conventional one by a factor 
	$n^{-\frac1p}$. This difference is motivated by the fact that in the 
	limit as $n \to \infty$, we can recover the continuous $L^p$ norm, formally.
	Define the $H^1_{per}(\bTd_n)$ norm of $f$ as follows:
		\begin{equation} \label{discreteh1norm}
		\|f\|_{H^1(\bTd_n)}:= \Big(\frac{1}{n} \sum_{x \in \bTd_n} n^2|f(x)-f(x-1)|
		^2 +  \frac{1}{n}\sum_{x \in \bTd_n} |f(x)|^2\Big)^\frac{1}{2}.
		\end{equation} 
 	We can also define the space $\ell^p(\mathbb{Z})$, with the following norm: For $f: 	
 	\mathbb{Z} \to \mathbb{C}$ and $p \geq 1$ define:
	\begin{equation} \label{Zlpnorm}
	\|f\|_{\ell^p(\mathbb{Z})}= \Big(\sum_{x \in \mathbb{Z}} |f(x)|^p\Big)^\frac{1}{p}.
	\end{equation}
	As usual we have:  $\ell^p(\bZ)= \{ f:\mathbb{Z} \to \mathbb{C}| \|f\|
	_{\ell^p(\mathbb{Z})}<\infty\}$.
	We  denote the discrete gradient of $f:\mathbb{Z} \to \mathbb{C}$ by $G(f)$, 
	and define it as:
		\begin{equation} \label{discreteZgradient}
			G(f):=\frac12\sum_{x \in \mathbb{Z}} |f(x)-f(x-1)|^2. 
		\end{equation} 
	Note the difference between $G(f)$ and $G_n(f)$ in \eqref{grad1}, where we scale 
	the second definition by $n^2$ in order to get the continuous counterpart, formally.  
\\
\\
	The first version of the discrete  Gagliardo-Nirenberg inequality can be 
	recalled from Proposition 17.6 of \cite{CH} with a small modification: 
	For every $1<p\leq \infty$, let $\theta=\frac12-\frac{1}{p+1}$, 
	obviously $\theta \in (0,1)$, we have: 
	$\forall f \in \ell^p(\mathbb{Z}) \cap \ell^2(\mathbb{Z})$, 
	there exists a constant $C(p)$ such that:
		
		\begin{equation} \label{GNdiscretep}
		\|f\|_{\ell^{p+1}(\mathbb{Z})} \leq C(p)\left(\|f\|_{\ell^2(\mathbb{Z})}\right)^{1-\theta}
		\left(G(f)\right)^{\frac{\theta}{2}}.
		\end{equation} 
	In particular, for $p=3$, we have  $\forall f \in \ell^4(\mathbb{Z}) 
	\cap \ell^2(\mathbb{Z})$, there exists a constant $C$ such that:
	
	\begin{equation} \label{GNdiscrete3}
 	\|f\|_{\ell^{4}(\mathbb{Z})}^4 \leq C(\|f\|_{\ell^2(\mathbb{Z})})^{3}(G(f))^{\frac12}. 
	\end{equation}
	We can deduce the following lemma from the later, which is crucial for 
	our purposes.
	
	\begin{lemma} \label{lemmaGNdiscerteperiodic}
	Recall the definition of $G_n$  \eqref{grad1}, and   $\|.\|_{\ell^p(\bTd_n)}$
	\eqref{discretelpnorm}, for every $f:\bTd_n \to \mathbb{C}$,  
	there exist a constant $C$ independent of $n$ such that:
		\begin{equation} \label{GNdiscreteperiodic1}
			\|f\|_{\ell^4(\bTd_n)}^4 \leq C\big((\|f\|_{\ell^2(
			\bTd_n)})^3(G_n(f))^{\frac12}	
			+\|f\|_{\ell^2(\bTd_n)}^2\big),
		\end{equation}
	we write this inequality in this open form:
	\begin{equation} \label{GNdiscreteperiodic2}
		\frac1n\sum_{x \in \bTd_n} |f(x)|^4 \leq 
		C\Bigg(\Big(\frac1n \sum_{x\in 	
		\bTd_n} 
		\frac{n^2}{2}|f(x)-f(x-1)|^2\Big)^{\frac12}\Big(\frac{1}{n} \sum_{x
		\in \bTd_n} |f(x)|^2\Big)^{\frac32}+\Big(\frac1n\sum_{x \in \bTd_n}|
		f(x)|^2\Big)^2\Bigg).
	\end{equation}
	Usually we have $\frac{1}{n} \sum_{x \in \bTd_n} |f(x)|^2=m$, 
	hence, we have:
		\begin{equation} \label{GNdiscreteperiodicm}
		\frac1n\sum_{x \in \bTd_n} |f(x)|^4 \leq C\left(m^{\frac32}G_n(f)^{\frac12}+m^2
		\right).
		\end{equation}
	\end{lemma}
	\begin{proof}
	We prove this lemma by constructing a function $\tilde{f} \in \ell^4(\mathbb{Z}) \cap
	 \ell^2(\mathbb{Z})$, from $f$ as follows: Translate  $f$ such that $|f(x)|^2$ 
	 achieves its minimum at $x=n$. By this construction, we have 
	 $|f(n)|^2 \leq \frac1n\sum_{x \in \bTd_n}|f(x)|^2=m $. Define $\tilde{f}$ 
	 on $\mathbb{Z}$ as:

		\begin{equation} \label{tempfunction}
			\tilde{f}(x)=
			\begin{cases}
			f(x), \: \: \forall x \in \{1,\dots,n\},
			\\ f(n)(2-\frac{x}{n}), \: \: \: \forall x \in \{n+1,\dots,2n\},
			\\ f(n)(1+ \frac{x}{n}), \: \: \: \forall x \in \{-n,\dots,-1\},
			\\ f(n), \:  \: \: \: \text{if} \quad x=0,
			\\0 \: \: \: \text{otherwise}.
			\end{cases}
		\end{equation}
		By the definition of $\tilde{f}$, for every $p \geq 1$ we have:
		\begin{equation}\label{lptempfunction}
			\|\tilde{f}\|_{\ell^p(\mathbb{Z})}^p=
			\sum_{x \in \bTd_n} |f(x)|^p + |f(n)|^p 
			\sum_{x=0}^{n-1} 2(\frac{x}{n})^p +|f(n)|^p.
		\end{equation}
	By estimating $\frac{1}{n}\sum_{x=1}^{n} (\frac{x}{n})^p$ with its integral value, 
	we  have $c_1(p)>0$, $c_2(p)>0$ independent of $n$, such that:
		\begin{equation} \label{tempestimate1}
		\|f\|_{\ell^p(\bTd_n)}^p + c_1(p)|f(n)|^p \leq \frac{1}{n} 
		\|\tilde{f}\|	
		_{\ell^p(\mathbb{Z})}^p \leq \|f\|_{\ell^p(\bTd_n)}^p + 
		c_2(p)|f(n)|^p.
		\end{equation}
	Moreover, we can compute $G(\tilde{f})$:
		\begin{equation} \label{tempgradientestimate}
		\begin{split}
		G(\tilde{f})= &\sum_{x\in \mathbb{Z}} \frac12|\tilde{f}(x)-\tilde{f}(x-1)|^2= 		
		\frac12 \sum_{x=1}^n |f(x)-f(x-1)|^2 +\\
		&|f(n)|^2(\sum_{x=-n}^0 \frac{1}{2n^2}+\sum_{x=n+1}^{2n} \frac{1}{2n^2})=\frac{1}	
		{n}G_n(f)+\frac{1}{n}|f(n)|^2.
		\end{split}
		\end{equation}
%We can deduce the following estimate using \eqref{tempgradientestimate} and \eqref{tempestimate1}
%and the fact that $|f(n)|^2 \leq \sum_{x \in \mathbb{T}_n} |f(x)|^2$
	Since we fix $n$, by \eqref{tempestimate1} $\tilde{f} \in \ell^2(\mathbb{Z})\cap 	
	\ell^4(\mathbb{Z})$; therefore, we can apply the inequality \eqref{GNdiscrete3}. By
	using the fact that $|f(n)|^2 \leq \frac{1}{n}\sum_{x \in \bTd_n} 
	|f(x)|^2=\|f\|^2_{\ell^2(\bTd_n)}$, and  estimates \eqref{tempgradientestimate} 
	and \eqref{tempestimate1}, we get:
	\begin{equation} \label{GNdiscereteperiodic2}
	\begin{split}
	\|f\|^4_{\ell^4(\bTd_n)} 
	&\leq \frac{1}{n} \|\tilde{f}\|_{\ell^4(\mathbb{Z})} \leq 
	C\big(\frac{1}{n}\|\tilde{f}\|^2_{\ell^2({\mathbb{Z})}}\big)^\frac{3}{2}\
	\big(nG(\tilde{f}) \big)^\frac{1}{2} 	
	\\ & \leq C\big(\|f\|^2_{\ell^2(\bTd_n)}+c_2(2)|f(n)|^2\big)
	^\frac{3}{2} \big( G_n(f)+|f(n)|^2\big)^\frac{1}{2}
	\\& \leq C'\big(\|f\|^2_{\ell^2(\bTd_n)}\big)^{\frac32}
	\big(G_n(f)+|f(n)|^2\big)^
	\frac{1}{2} \\
	& \leq C'\big(\|f\|^2_{\ell^2(\bTd_n)}\big)^{\frac32}G_n(f)^{\frac12} 
	+C'\big(\|f\|^2_{\ell^2(\bTd_n)}\big)^{\frac32}\|f\|_{\ell^2
	(\bTd_n)}^4.
	\end{split}
	\end{equation}
	This inequality proves the lemma, since $C'=C(1+c_2(2))^{\frac32}$ 
	is a constant independent of $n$. In the last line, we used the inequality 
	$\sqrt{a+b} \leq \sqrt{a}+\sqrt{b}$ for $a,b>0$. 
      \end{proof}

		\section{Jacobi Elliptic Functions} \label{JEF}
	   Given $k \in (0,1)$, the incomplete elliptic integral of the first kind,
	   for any $\phi \in \bR$ is defined as: 
	   $$x= F(\phi;k):= \int_0^{\phi} \frac{d \theta}{\sqrt{1-k^2 \sin^2(\theta)}}.$$
	   Consequently, one can define $\cn(\cdot),\sn(\cdot),\dn(\cdot)$ via the inverse
	   of $F(\cdot,k)$:
	   \begin{equation} \label{Jacobielliptic}
	   \sn(x,k):= \sin(\phi), \quad \cn(x,k):= \cos(\phi), \quad \dn(x,k):= 
	   \sqrt{1-k^2 \sin^2(\phi)}.
	   \end{equation}
		From \eqref{Jacobielliptic}, it is straightforward to see for all $x$ 
		\begin{equation}
			\sn^2(x,k)+\cn^2(x,k)=k^2\sn^2(x,k)+\dn^2(x,k)=1.
		\end{equation}
		Moreover, the derivative (w.r.t $x$) 
		of these functions can be obtained directly from
		the definition:
		\begin{equation} \label{Jacobiderivative}
			\begin{split}			
			\partial_x \sn(x,k) =\cn(x,k) \dn(x,k), \quad \partial_x \cn(x,k)=-\sn(x,k)
			\dn(x,k), \quad  \\\partial_x \dn(x,k) = -k^2 \cn(x,k) \sn(x,k). 					
			\end{split}		
		\end{equation}
		Moreover, the period of these functions is given via the following complete 
		elliptic integral:
		\begin{equation}
			K(k):= F(\frac{\pi}{2};k), 
		\end{equation}
		where $\dn$ is $2K$ periodic and even, $\sn$ and $\cn$ are $4K$ periodic, where
		$\sn$ is $2K$ anti periodic and odd, and $\cn$ is $2K$ anti periodic and even. \\
		Notice the limiting cases: $K(k) \to \frac{\pi}{2}$ as $k \to 0$, and $K(k) \to
		 \infty$ as $k \to 1$. Moreover, as for $k = 0$, $\sn(x,0)=\sin(x) $, 
		$\cn(x,0)=\cos(x)$, $\dn(x,0)=1$. Furthermore, $\sn(x,1)=\tanh(x)$, 
		$\cn(x,1)=\dn(x,1)=\sech(x)$.\\
		Finally, notice that from \eqref{Jacobiderivative} one can deduce that 
		$\frac{1}{\alpha}\dn(\frac{x}{\beta},k)$, 
		$\frac{1}{\alpha}\cn(\frac{x}{\beta},k)$, and 
		$\frac{1}{\alpha}\sn(\frac{x}{\beta},k)$ are solutions to \eqref{ODE1}, 
		where $\alpha,\beta,k$ are determined by $\omega$, $L$ in each case, respectively. 

\end{appendices} 

\bibliographystyle{plain}
\bibliography{bibilo}

\end{document}